\documentclass[aps,prx,twocolumn,superscriptaddress,showpacs,floatfix]{revtex4-2}

\usepackage{amsmath,amssymb}
\usepackage{newtxtext,newtxmath}
\usepackage[scaled=0.92]{helvet}

\usepackage{amsfonts,float,color,verbatim,tabularx,bm,appendix,url}
\usepackage{xcolor}
\usepackage{graphicx}
\usepackage{multirow}
\usepackage{booktabs}
\usepackage[colorlinks=true, citecolor=blue, linkcolor=blue, urlcolor=blue, breaklinks=true]{hyperref}
\usepackage{setspace}
\usepackage{hyperref}
\usepackage{orcidlink}
\usepackage{soul}

\begin{document}

\title{Numerical evidence of a critical point in the (2+1)D SO(5) nonlinear sigma model with Wess-Zumino-Witten term}

\author{Yuan Da Liao \orcidlink{0000-0002-3018-4029}}
\affiliation{Department of Physics and HK Institute of Quantum Science \& Technology, The University of Hong Kong, Pokfulam Road, Hong Kong SAR, China}
\affiliation{State Key Laboratory of Optical Quantum Materials, The University of Hong Kong, Pokfulam Road, Hong Kong SAR, China}

\author{Bin-Bin Chen}
\affiliation{Peng Huanwu Collaborative Center for Research and Education, Beihang University, Beijing 100191, China}

\author{Junchen Rong\orcidlink{0000-0001-5749-1574}}
\affiliation{CPHT, CNRS, \'Ecole Polytechnique, Institut Polytechnique de Paris, Palaiseau, France}

\author{Fakher F. Assaad \orcidlink{0000-0002-3302-9243}}
\affiliation{Institut f\"ur Theoretische Physik und Astrophysik, Universit\"at W\"urzburg, 97074 W\"urzburg, Germany}
\affiliation{W\"urzburg-Dresden Cluster of Excellence ctd.qmat, Am Hubland, 97074 W\"urzburg, Germany}

\author{Lukas Janssen \orcidlink{0000-0003-4919-796X}}
\affiliation{Institut f\"ur Theoretische Physik and W\"urzburg-Dresden Cluster of Excellence ctd.qmat, TU Dresden, 01062 Dresden, Germany}

\author{Zi Yang Meng \orcidlink{0000-0001-9771-7494}}
\affiliation{Department of Physics and HK Institute of Quantum Science \& Technology, The University of Hong Kong, Pokfulam Road, Hong Kong SAR, China}
\affiliation{State Key Laboratory of Optical Quantum Materials, The University of Hong Kong, Pokfulam Road, Hong Kong SAR, China}

\begin{abstract}
We develop an optimized continuous-field quantum Monte Carlo (QMC) algorithm to investigate the projected SO(5) nonlinear sigma model with a Wess-Zumino-Witten term, which describes half-filled Dirac fermions in 2+1 space-time dimensions akin to graphene and Yukawa coupled to a quintuplet of compatible mass terms. 
Our algorithm reduces the computational complexity to $O(\beta N_{\mathbf{q}} N_\phi^2)$, yielding a speedup of a factor of $N_\phi$ (the number of magnetic fluxes, i.e., system size) relative to prior works~\cite{ippolitiHalf2018,wangPhases2021,chenPhases2023,huffman2026generalizingdeconfinedcriticality3d}. 
This advance enables us to simulate system sizes up to $N_\phi=140$ on the torus and $N_\phi=59$ on the sphere, far exceeding the maximum sizes previously accessed, and to map out the universal phase diagram of the model on both geometries. 
Most notably, we identify and characterize a critical point that separates an SO(5)-broken ordered phase at small coupling from an SO(5)-symmetric disordered phase at large coupling. 
The critical point becomes multicritical upon the inclusion of terms that break the SO(5) symmetry down to $\mathrm{U}(1) \times \mathrm{SU}(2)$, relevant for the deconfined phase transition between N\'eel antiferromagnetic and valence-bond-solid orders in quantum magnets. 
Our finding of a multicritical point in the phase diagram of the SO(5) nonlinear sigma model with Wess-Zumino-Witten term resolves the long-standing open question of its global structure, and our QMC algorithm opens a new avenue for systematic studies of projected Hamiltonians, ranging from correlated flat bands to fractional quantum (anomalous) Hall systems.
\end{abstract}

\date{\today}
\maketitle

\section{Introduction}
\label{sec:introduction}
When Dirac fermions are subjected to a perpendicular magnetic field, they form discrete, highly degenerate Landau levels~\cite{zhengHall2002,novoselov2005,zhangExperimental2005,liObservation2007}. 
For half-filled Dirac fermions, projecting a Yukawa coupling to a quintuplet of compatible mass terms onto the lowest Landau level (LLL) and integrating out high-energy fermionic degrees of freedom, generates a Wess-Zumino-Witten topological term in the bosonic action of fermion bilinears~\cite{leeWess2015}. 
The resulting effective low-energy theory is a (2+1)D SO(5) nonlinear sigma model with a level-1 Wess-Zumino-Witten term, which hosts both Néel antiferromagnetic (AFM) and Kekulé valence-bond solid (VBS) phases as symmetry-broken ground states, as well as possibly continuous phase transitions between them~\cite{haldaneFractional1983,affleckCritical1987,thetaAbanov2000,tanakaBerry2005,senthilCompeting2006,leeWess2015,ippolitiHalf2018,wangPhases2021,zhouThe2024,chenPhases2023,huffman2026generalizingdeconfinedcriticality3d}. 
This model is intrinsically linked to the paradigm of deconfined quantum criticality, which describes continuous phase transitions between phases with incompatible broken symmetries beyond the Landau-Ginzburg-Wilson framework~\cite{Harada2003,senthilQuantum2004,sandvikEvidence2007,qinDuality2017,wangDeconfined2017,maDynamics2018,liuSuperconductivity2019,maRole2019,senthilDeconfined2023}.

Recent numerical studies of deconfined quantum criticality in lattice spin models (e.g., the $J$-$Q$ model~\cite{sandvikEvidence2007})
have raised critical questions about the nature of the transition, with accumulating evidence suggesting 
a weakly-first-order transition
rather than a generic continuous quantum critical point~\cite{kuklovDeconfined2008,jiangFrom2008,chenDeconfined2013,nahumDeconfined2015,blockFate2013,haradaPossibility2013,takahashi2024so5}. In particular, entanglement entropy measurements have revealed anomalous logarithmic corrections at the AFM-to-VBS transition points, which have been attributed to residual Goldstone modes associated with weak-first-order behavior~\cite{zhaoScaling2022,wangScaling2022,liuFermion2023,liaoTeaching2023,zhaoScaling2024,songExtracting2023,song2023deconfined,emidioEntanglement2024,zhaoScaling2025,zhuBipartite2026}.
The weakly-first-order nature of the transition may be due to a fixed-point annihilation mechanism~\cite{nahumDeconfined2015, gorbenko2018, nahumNote2020, wangDeconfined2017, maTheory2020, zou21, He:2021xvg, hawashin25, ray25} or the presence of a nearby multicritical point~\cite{chesterBootstrapping2024, takahashi2024so5} with a slightly relevant operator.

\begin{figure}[tp!]
\includegraphics[width=\columnwidth]{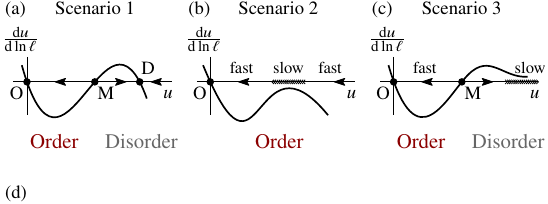}\\
\includegraphics[width=\columnwidth]{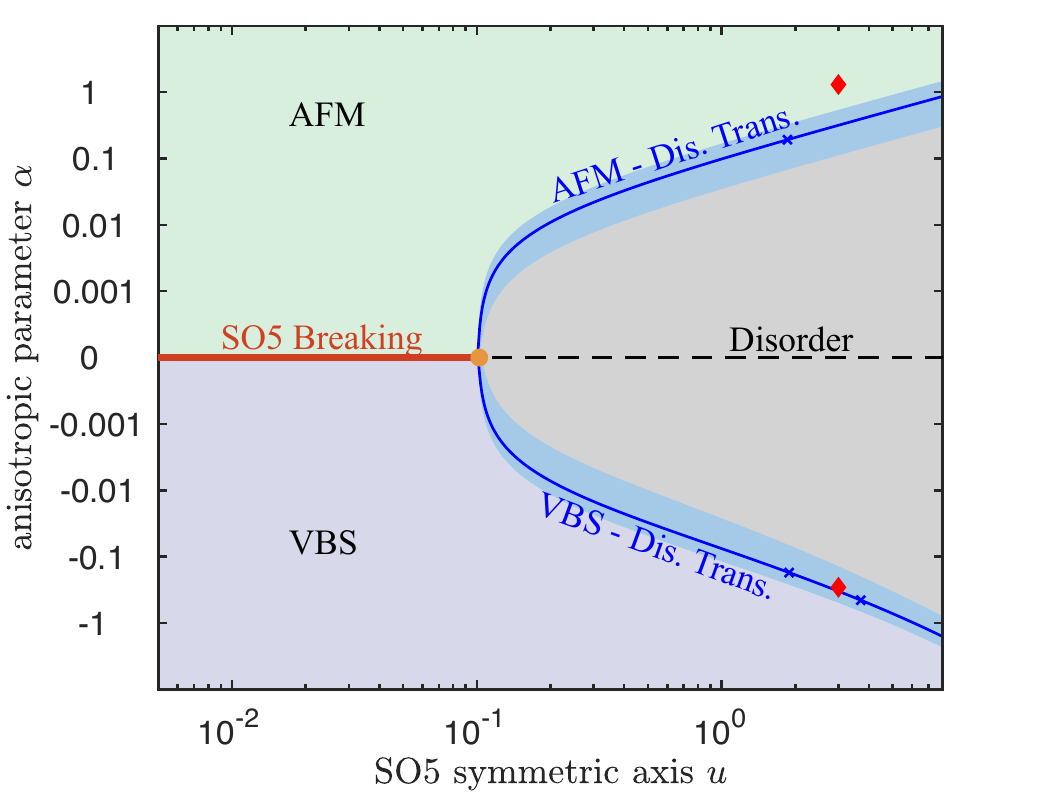}
\caption{\textbf{Possible renormalization group flows of the nonlinear sigma model with Wess-Zumino-Witten term and universal phase diagram.}
(a)~Scenario~1: For sufficiently large $N$, the SO($N$) nonlinear sigma model has an infrared stable noninteracting fixed point `O', describing an SO($N$) ordered phase, a stable interacting fixed point `D', describing a gapless disordered phase, and a critical fixed point `M' in between. Arrows denote flow towards the infrared.
(b)~Scenario~2: For decreasing values of $N$, fixed point `D' may move towards and eventually annihilate with the critical fixed point `M', resulting in an instability of the disordered state towards the long-range ordered state.
(c)~Scenario~3: Alternatively, fixed point `D' may move away from `M' upon decreasing $N$, eventually either annihilating with another fixed point at finite coupling (not shown) or approaching the strong-coupling fixed point.
(d)~Phase diagram of the model in Eq.~\eqref{eq:hamiltonian_regularized_main} in the parameter space spanned by the SO(5) coupling $u$ and the anisotropy $\alpha$. The diagram contains four regimes: an SO(5)-symmetry-broken phase at small $u$ (red segment), an SO(5)-symmetric disordered phase with chiral-spin-liquid-like character (grey), a Néel AFM phase for $\alpha>0$ (green), and a Kekulé VBS phase for $\alpha<0$ (cyan). The orange point indicates the multicritical point separating the ordered SO(5)-breaking phases from the disordered phase. The blue curves and shaded bands show the phase boundaries and associated uncertainties obtained in the previous DMRG study on the spherical geometry. Red diamonds denote the transition points extracted from our QMC simulations. QMC error bars are omitted for clarity, since the dominant uncertainties arise from systematic finite-size-scaling effects away from the SO(5)-symmetric line; conservative estimates can be broad and may, for some cuts, approach or include $\alpha=0$ (see Sec.~\ref{sec:results_phaseboundaries}).
}
\label{fig:phasediagram}
\end{figure}

Both of these two seemingly competing scenarios can be understood within the framework provided by the SO(5) nonlinear sigma model with Wess-Zumino-Witten term, which represents a minimal continuum description for the deconfined phase transition. 
The model belongs to a family of nonlinear sigma models with the target space being the Stiefel-manifold SO($N$)/SO(4), which are amenable to a $1/N$ expansion~\cite{zou21, ray25}. 
The model has an explicit SO($N$)$\times$SO($N-$4) symmetry.
For sufficiently large Wess-Zumino level $k$, these models support three fixed points, see Fig.~\ref{fig:phasediagram}(a) (``Scenario~1''). The noninteracting fixed point, marked as `O', describes a stable ordered phase, while the interacting stable fixed point `D' realizes a gapless disordered phase, described by a unitary conformal field theory. 
The two phases are separated by the critical fixed point marked as `M' in Fig.~\ref{fig:phasediagram}(a), which is described by another unitary conformal field theory. The critical fixed point `M' becomes multicritical upon the inclusion of explicit SO($N$)$\times$SO($N-$4)-symmetry-breaking terms.

The fate of the interacting fixed points in the most physically relevant case of $N=5$ is less obvious.
Previous theoretical works~\cite{nahumNote2020, maTheory2020, zou21, He:2021xvg, hawashin25, ray25} have entertained the possibility that the interacting stable fixed point `D' moves towards and eventually annihilates with the critical fixed point `M' upon decreasing $N$ beyond a critical $N_\mathrm{c} > 5$, resulting in an instability of the interacting disordered state towards the long-range ordered state, see Fig.~\ref{fig:phasediagram}(b) (``Scenario~2'').
In this scenario, the ground state of the SO(5) nonlinear sigma model with Wess-Zumino-Witten term spontaneously breaks SO(5) symmetry throughout the phase diagram. However, the fixed-point annihilation results in a slow flow near the complexified fixed points~\cite{wangDeconfined2017} and an exponential suppression of the order parameter $\sim \mathrm e^{-a/\sqrt{N_\mathrm{c}-N}}$, with $a>0$ a nonuniversal constant, in the strong-coupling regime.

An alternative scenario is that the fixed point `D' moves away from `M' upon decreasing $N$, eventually either annihilating with another fixed point at finite coupling (not shown) or approaching the strong-coupling fixed point, see Fig.~\ref{fig:phasediagram}(c) (``Scenario~3'').
Independent of the fate of the fixed point `D', the critical point `M' in this scenario is not involved in any fixed-point collision all the way down to $N=5$ and remains described by a unitary conformal field theory. 
It separates the SO(5)-ordered phase at small coupling from an SO(5)-disordered phase at large coupling. 
The disordered phase has to be either gapless or spontaneously break some other symmetry, due to the nontrivial 't Hooft anomaly~\cite{zou21}. 
Notable candidates include the chiral spin liquid phase as proposed in~\cite{wangDeconfined2017} and the $Z_2$ gapless spin liquid proposed in~~\cite{Shackleton:2021fdh}. Determining which scenario is realized for $N=5$ requires precise numerical determination of the phase structure and the operator scaling dimensions at scale-invariant points. 
In this work, we provide robust numerical evidence on the existence of the critical point `M', through systematic finite-size scaling analysis on unprecedented system sizes, extracting the critical exponents $\Delta_\varv = 0.52(1)$ and 
$\Delta_s = 2.2(6)$ on the spherical geometry, with consistent results on the torus.

The SO(5) nonlinear sigma model with a Wess-Zumino-Witten term has been studied numerically with two distinct regularization schemes: toric geometry with periodic boundary conditions~\cite{ippolitiHalf2018,wangPhases2021}, and spherical geometry with a magnetic monopole, which preserves full rotational symmetry and eliminates edge effects~\cite{zhouThe2024,chenPhases2023,chenEmergent2024,huffman2026generalizingdeconfinedcriticality3d}. 
The study of LLL models on spherical geometry allows one to take advantage of the state-operator correspondence of conformal field theory. 
By looking for integer-spaced conformal towers, one can detect critical points with a small system size. 
This led the authors of~\cite{zhuUncovering2023} to initiate the ``fuzzy sphere'' program. 
When applied to the SO(5) Wess-Zumino-Witten model, they observed approximate conformal towers~\cite{zhouThe2024} possibly due to ``pseudo-critical'' behavior in Scenario 2.
It is, however, important to check the validity of this scenario at larger system size, which is the goal of this paper. 
Ref.~\cite{chenPhases2023} has explored the SO(5) Wess-Zumino-Witten model with deformations breaking the SO(5) symmetry, and reported critical points separating an SO(5)-symmetric phase with finite width from the VBS and AFM phases. 
This raises the possibility of a stable, extended disordered SO(5)-symmetric phase in the spirit of a quantum spin liquid in the phase diagram, consistent with the structure of the phase diagram proposed in the generalized $J$-$Q$ model~\cite{takahashi2024so5}.
The exact position of the multicritical point suffers from sizable finite-size effects, and one therefore needs to find alternative tools to access the converged phase boundaries in the thermodynamic limit.
Previous studies have been severely limited by the computational bottlenecks of conventional discrete auxiliary-field determinantal quantum Monte Carlo (QMC) method~\cite{ippolitiHalf2018,wangPhases2021,chenPhases2023,huffman2026generalizingdeconfinedcriticality3d}, not to mention the even smaller system sizes in exact diagonalization and density-matrix renormalization group (DMRG) methods~\cite{zhouThe2024,chenEmergent2024}. 
Thus, advancing numerical algorithms is imperative. 

{We design a QMC method} with significantly reduced computational complexity to study the phase diagram of the SO(5) nonlinear sigma model with Wess-Zumino-Witten term, {on both torus and sphere geometries}. Our advance is based on the observation that, different from the discrete-auxiliary-field QMC commonly used~\cite{ippolitiHalf2018,wangPhases2021,chenPhases2023,huffman2026generalizingdeconfinedcriticality3d}, one can follow the recent development in the momentum-space QMC for twisted bilayer graphene~\cite{angleHuang2024,huangStrain2026,shanTrion2026} (also a projection problem with long-range interaction and regularization) to use a continuous auxiliary field in the QMC simulations. With the continuous auxiliary field, we can perform a global update for an entire imaginary time slice, i.e., $N_\phi$ sites get updated all together, and use  Langevin dynamics~\cite{langevinBatrouni1985,langevinBatrouni2019,valenceGeotz2022} supplemented by a Metropolis 
acceptance criterion~\cite{robertsOptimal1998,robertsOptimal2001}. 
Our improved continuous field determinant QMC (CF-DQMC) algorithm is shown to be $O(N_\phi)$ faster than the previous simulations on torus and sphere geometries~\cite{ippolitiHalf2018,wangPhases2021,chenPhases2023,huffman2026generalizingdeconfinedcriticality3d}.
With such improvement and the new understanding of the phase diagram of the problem~\cite{chenPhases2023}, in this work we (1)~successfully simulated system sizes with $N_\phi=140$ on the torus and $N_\phi=59$ on the sphere, significantly larger than the previous maximum of $N_\phi=100$ on the torus~\cite{wangPhases2021} and $N_\phi=21$~\cite{huffman2026generalizingdeconfinedcriticality3d} on the sphere, and determined the precise critical exponents of the critical point, and (2)~explored a much larger parameter space than previous works, with ease and significantly less computational cost. 
The obtained phase diagram is shown in Fig.~\ref{fig:phasediagram}(d).

These simulations reveal that the triple point that separates the AFM and VBS ordered phases from the SO(5)-disordered phase (orange dot in Fig.~\ref{fig:phasediagram}(d)) is a multicritical point that is fully compatible with a unitary conformal-field-theory description, including the most recent bounds from conformal bootstrap calculations~\cite{nakayamaNecessary2016, polandConformal2019, li22,chesterBootstrapping2024}. 
The presence of the critical point along the SO(5) line refutes the fixed-point annihilation Scenario~2 outlined in Fig.~\ref{fig:phasediagram}(b).
On both spherical and toric geometries, our numerical results are fully compatible with a single interacting fixed point `M' at finite coupling.
This bears a striking similarity to that of the generalized $J$-$Q$ model~\cite{takahashi2024so5}. 
In particular, a first-order transition across the SO(5)-breaking phase, together with a multicritical point and a disordered (spin-liquid) phase, separates the N\'eel and VBS phases.
Our numerical results in the disordered phase show that the current maximal system size is not yet able to determine the precise nature of the phase. 
Whether Scenario~1 or Scenario~3 is actually realized is therefore left for future work.

\section{Results}
\label{sec:results}

\subsection{Theoretical model, Fierz identity transformation, and computational bottleneck}
\label{sec:results_model}
Our starting point for numerical simulation is the regularized momentum-space Hamiltonian for the projected SO(5) model~\cite{ippolitiHalf2018, wangPhases2021, chenPhases2023}. By projecting the continuum field operator onto the LLL basis, we obtain
\begin{equation}
\mathcal{H} = U_0 \mathcal{H}_0 -\sum_{i=1}^5 U_i \mathcal{H}_i,
\label{eq:hamiltonian_regularized_main}
\end{equation}
where
\begin{equation}
\mathcal{H}_i=\sum_{\mathbf{q}}^{N_{\mathbf{q}}} n^{\Gamma^{i}}_{\mathbf{q}} n^{\Gamma^{i}}_{-\mathbf{q}}.
\end{equation}
Here, $n^{\Gamma^{i}}_{\mathbf{q}} = \int d\boldsymbol{r} e^{i\mathbf{q}\cdot\boldsymbol{r}} \hat{\boldsymbol{\psi}}^{\dagger}_{\boldsymbol{r}} \Gamma^i \hat{\boldsymbol{\psi}}_{\boldsymbol{r}}$ is the Fourier-transformed density operator for the $i$-th SO(5) component, with $\hat{\boldsymbol{\psi}}^{\dagger}_{\boldsymbol{r}}$ ($\hat{\boldsymbol{\psi}}_{\boldsymbol{r}}$) written in different basis functions on torus and sphere geometries (see Sec.~\ref{sec:methods_continuum} in Methods) and $\Gamma^i$ being the 4×4 Gamma matrices encoding the valley and spin degrees of freedom. 
The $4\times4$ Gamma matrices
\begin{equation}
\Gamma^i=\{\tau_x\otimes\mathbb{I}_2, \tau_y\otimes\mathbb{I}_2, \tau_z\otimes{\sigma_x}, \tau_z\otimes{\sigma_y}, \tau_z\otimes{\sigma_z}\}
\end{equation}
are mutually anticommuting, and their commutators $L^{ij}=\tfrac{-i}{2}[\Gamma^i,\Gamma^j]$ form the 10 generators of the SO(5) Lie group.We also define the $0$-th component $\Gamma^0$ as the identity operator. More details of the model can be found in Sec.~\ref{sec:methods_fierz} of Methods. 
We set $U_0=1$ as the energy unit, and parameterize the interaction strengths as $U_1=U_2=U_K$ (coupling to valley/Kekulé VBS order) and $U_3=U_4=U_5=U_N$ (coupling to spin/Néel order). When $U_K=U_N=u>0$, the Hamiltonian has exact SO(5) symmetry.

A key first step in our algorithm is to perform a Fierz identity transformation on the Hamiltonian in Eq.~\eqref{eq:hamiltonian_regularized_main}. This transformation reorganizes the quartic fermionic interaction terms by exchanging spin indices, resulting in the transformed Hamiltonian
\begin{equation}\label{eq:hamiltonian_transformed_main}
    H=\sum_{i=0}^{3} g_i H_i,
\end{equation}
where
\begin{equation}
H_i =\sum_{\mathbf{q}}^{N_{\mathbf{q}}} n^{O^{i}}_{\mathbf{q}} n^{O^{i}}_{-\mathbf{q}}.
\end{equation}
The transformed operators are 
\begin{equation}
O^i=\{\mathbb{I}_2\otimes\mathbb{I}_2,\tau_x\otimes\mathbb{I}_2, \tau_y\otimes\mathbb{I}_2, \tau_z\otimes\mathbb{I}_2\},
\end{equation}
and the renormalized coupling constants are $g_0 = U_0+U_N$, $g_1=-(U_K+U_N)$, $g_2=-(U_K+U_N)$, and $g_3=2U_N$.

This Fierz identity transformation provides two critical advantages:
(1) Complete elimination of the fermionic sign problem in the presence of only contact interactions, since the transformed Hamiltonian has fully decoupled spin degrees of freedom, allowing us to exploit the particle-hole symmetry of the half-filled system to ensure the fermion determinant is positive definite for all parameter values studied.
(2) The number of independent interaction channels is reduced from 6 to 4, from $\{\Gamma^0, \Gamma^1,\cdots,\Gamma^5\}$ to $\{O^0, O^1, \cdots,O^3\}$, simplifying the subsequent Hubbard-Stratonovich (HS) decomposition and reducing the number of auxiliary fields required in the Monte Carlo sampling.
Full details of the model regularization and Fierz identity transformation are provided in Sec.~\ref{sec:methods_fierz} in the Methods.

In Eqs.~\eqref{eq:hamiltonian_regularized_main} and \eqref{eq:hamiltonian_transformed_main}, the system size $N_\phi$ enters from the number of orbits in the basis  $\hat{\boldsymbol{\psi}}^{\dagger}_{\boldsymbol{r}}$ ($\hat{\boldsymbol{\psi}}_{\boldsymbol{r}}$) and denotes the degeneracy of the LLL, which is equal to the number of magnetic flux quanta piercing the system and serves as our measure of the volume. $N_{\mathbf{q}}$ denotes the number of momentum points included in the simulation. For torus geometry, a natural momentum cutoff arises from the Gaussian form factor of the LLL wave functions, giving $N_{\mathbf{q}} \propto N_\phi$. For sphere geometry, we include all angular momentum channels, giving $N_{\mathbf{q}}=(N_\phi+1)^2 \propto N_\phi^2$, which renders a higher computational complexity than that of the torus.

Conventional discrete auxiliary-field determinant QMC simulations of this model face a severe computational bottleneck. The projected interactions are long-ranged, a local update of a single auxiliary field leads to a non-local change in the fermion determinant, requiring $O(N_\phi^3)$ operations per update (dominated by matrix inversion and multiplication). Since $O(N_{\mathbf{q}})$ auxiliary fields need to be updated per imaginary time slice, and there are $O(\beta/\Delta\tau)$ time slices (where $\beta$ is the inverse temperature and $\Delta\tau$ is the Trotter step size), the overall computational complexity is $O(\beta N_{\mathbf{q}} N_\phi^3)$.
For the spherical geometry, where $N_{\mathbf{q}}\propto N_\phi^2$, this complexity rises to $O(\beta N_\phi^5)$ and it is $O(\beta N_\phi^4)$ on torus, which has limited prior simulations to $N_\phi\leq 21$ on sphere~\cite{chenPhases2023,huffman2026generalizingdeconfinedcriticality3d} and $N_\phi\leq 100$ on the torus~\cite{ippolitiHalf2018,wangPhases2021}.

\begin{figure*}[htp!]
\includegraphics[width=\textwidth]{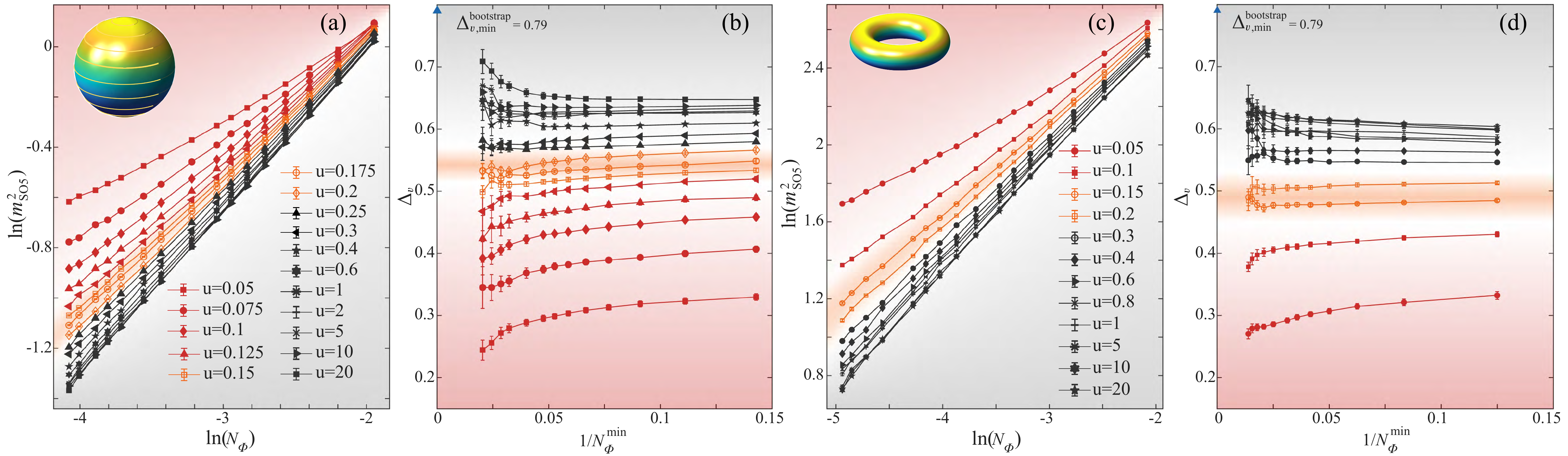}
\caption{\textbf{SO(5) order parameter scaling on sphere and torus geometries.} (a,b) Results for the sphere geometry with $N_\phi$ up to 59; (c,d) results for the torus geometry with $N_\phi$ up to 140. (a,c) Log-log plot of the SO(5) order parameter $m_{\text{SO(5)}}^2$ versus system size $N_\phi$ for different values of $u$ along the SO(5)-symmetric line. (b,d) Scaling dimension $\Delta_\varv$ of the SO(5) order parameter versus inverse system size $1/N_\phi^{\text{min}}$, where $N_\phi^{\text{min}}$ is the minimum system size included in the scaling fit. Three distinct regimes are observed: a SO(5) symmetry-broken phase (red shaded area, $u<0.15$), a multicritical point (orange shaded area, $u \sim 0.18$), and a stable SO(5) symmetric disordered phase (grey shaded area, $u>0.2$). The phase structure is consistent across both geometries, as in the universal phase diagram in Fig.~\ref{fig:phasediagram} (d). $\Delta_{\varv,\mathrm{min}}^{\mathrm{bootstrap}}$ denotes the conformal bootstrap lower bound on the scaling dimension of the SO(5) vector in an infrared-stable conformal phase~\cite{li22}.
} 
\label{fig:SO5line}
\end{figure*}

\subsection{Continuous field QMC with MALA global updates}
\label{sec:results_mala}
To overcome the computational bottleneck of conventional determinant QMC, we develop an optimized continuous field determinant QMC algorithm with a Metropolis-adjusted Langevin algorithm (MALA) for global updates of continuous auxiliary fields. This approach is enabled by our Fierz-transformed Hamiltonian and a continuous Hubbard-Stratonovich (HF) decomposition, as detailed in Sec.~\ref{sec:methods_hs} in Methods.

The key insight is that, unlike discrete auxiliary fields which require point-by-point local updates, continuous auxiliary fields allow us to propose updates for all $N_{\mathbf{q}}$ auxiliary fields in an entire imaginary time slice simultaneously. To implement these global updates efficiently with a high acceptance rate, we use the MALA~\cite{robertsOptimal1998,robertsOptimal2001} technique, which combines Langevin dynamics with a Metropolis accept/reject step to ensure detailed balance.

The global update proceeds as follows. For a given imaginary time slice $\tau$, we propose a new set of auxiliary field configurations $\{\phi_{\tau,q}^{'i}\}$ from the current configurations $\{\phi_{\tau,q}^i\}$ for the entire time slice, via the Langevin dynamics equation:
\begin{equation}
    \phi_{\tau,q}^{i\ \prime} = \phi_{\tau,q}^i + \frac{\epsilon^2}{2}  \nabla_{\phi_{\tau,q}^i} S[\phi] +\epsilon \mathcal{N}(0,1),
\label{eq:langevin_proposal_main}
\end{equation}
where $\epsilon$ is the Langevin step size, $\mathcal{N}(0,1)$ is the standard normal distribution, and $S[\phi]$ is the total effective action of the auxiliary fields based on the equal time single-particle Green's function and the imaginary time propagator constructed from the auxiliary field configuration. The critical ingredient is the gradient of the action, $\nabla_{\phi_{\tau,q}^i} S[\phi]$, which can be computed directly from the trace operation of the multiplication of equal-time Green's functions with a computational complexity of $O(N_\phi^2)$ per update. The deviation of the gradient is given in Sec.~\ref{sec:methods_langevin} in Methods.

The proposed configuration $\{\phi'\}$ is then accepted or rejected according to the Metropolis-Hastings criterion. To optimize the update efficiency, we implement an adaptive step size adjustment mechanism that automatically tunes $\epsilon$ during the simulation to maintain an average acceptance rate of $r_0=0.574$, the theoretically optimal value for MALA~\cite{robertsOptimal1998,robertsOptimal2001}.

The overall computational complexity is thus reduced to $O(\beta N_{\mathbf{q}} N_\phi^2)$,
which represents a speedup of a factor of $N_{\phi}$ relative to prior methods with $O(\beta N_{\mathbf{q}} N_\phi^3)$. 
This acceleration enables us to readily simulate unprecedented system sizes up to $N_\phi=140$ on torus and $N_\phi=59$ on sphere, far beyond previous limits (for example, Ref.~\cite{huffman2026generalizingdeconfinedcriticality3d} only simulate $N_\phi=29$). We have benchmarked our algorithm against DMRG results for small system sizes, finding perfect agreement for all physical observables (see Sec.~\ref{sec:methods_benchmark} in the Methods), confirming that our method is unbiased and accurate.

\subsection{Universal phase diagram on two geometries}
\label{sec:results_phase_diagram}
We first present the universal phase diagram of the projected SO(5) model on sphere and torus geometries as our central finding, visualized in Fig.~\ref{fig:phasediagram} (d). This phase diagram is parameterized by two tuning parameters: the horizontal axis is $u$, which defines the SO(5)-symmetric line where the interaction strengths satisfy $U_K=U_N=u$; the vertical axis is the anisotropy parameter $\alpha$, defined via the relations $U_N = u+\alpha/\sqrt{2}$ and $U_K = u - \alpha/\sqrt{2}$, which breaks the full SO(5) symmetry down to $\text{SO(3)}\times\text{SO(2)}$ for non-zero $\alpha$.

The phase diagram exhibits a universal structure consisting of four distinct thermodynamic phases and associated phase transitions. At small values of $u \lesssim 0.2$ along the SO(5)-symmetric line, the system resides in a phase with spontaneously broken SO(5) symmetry (red segment, small $u$ in Fig.~\ref{fig:phasediagram} (d)). This symmetry-broken phase is separated from a SO(5)-symmetric disordered phase by a multicritical point located near $u \sim 0.2$ (orange dot in Fig.~\ref{fig:phasediagram} (d)). For all values of $u > 0.2$, the system remains in this SO(5)-symmetric disordered state
separates the two symmetry-broken phases away from the SO(5)-symmetric line. For positive values of $\alpha$, where the Néel coupling $U_N$ dominates over the valence-bond solid (VBS) coupling $U_K$, the system forms a Néel antiferromagnetic (AFM) phase with spontaneous breaking of the SO(3) spin symmetry (green area in Fig.~\ref{fig:phasediagram} (d)). Conversely, for negative values of $\alpha$, where the VBS coupling $U_K$ is dominant, the system enters a Kekulé VBS phase with spontaneous breaking of the SO(2) valley symmetry (cyan area in Fig.~\ref{fig:phasediagram} (d)).

Continuous phase transitions between the symmetric disordered phase and the AFM/VBS phases are indicated by blue lines and crosses in Fig.~\ref{fig:phasediagram} (d), with the blue shaded region denoting the uncertainty in the determined critical points.  
The red diamonds and blue crosses mark the transition points determined in this work and in a previous density matrix renormalization group study on sphere geometry~\cite{chenPhases2023}, respectively. 
Interestingly, the overall topological structure of the phase diagram is consistent across both spherical and toric geometries, and even the position of the multicritical point arising from the different regularization schemes is similar, as discussed below. This consistency confirms that the observed phase structure is a universal feature of the model in the thermodynamic limit, rather than a finite-size or geometry-specific artifact.

\subsection{Phase structure along the SO(5)-symmetric line}
\label{sec:results_so5line}
We now detail the phase behaviour along the SO(5)-symmetric line for two geometries, which constitutes the horizontal axis of the phase diagram presented in Fig.~\ref{fig:phasediagram} (d). To characterize the phase structure, we use the SO(5) order parameter defined as
\begin{equation}
m_{\text{SO(5)}}^2 = \frac{1}{5N_{\phi}^2}\sum_{i=1}^5 S^i_{\mathbf{q}=0},
\end{equation}
where $S^i_{\mathbf{q}} = \langle n_{\mathbf{q}}^{\Gamma^i} n_{\mathbf{-q}}^{\Gamma^i} \rangle$ denotes the static correlation function for the $i$-th SO(5) component. We further extract the effective scaling dimensions $\Delta_\varv$ of the SO(5) order parameter via finite-size scaling analysis, using the power-law relation $m_{\text{SO(5)}}^2 \sim N_\phi^{-\Delta_{\varv}}$. Since our continuous field QMC performs at its best along the SO(5) line, we performed the simulation with $N_\phi=140$ on torus geometry and $N_\phi=59$ on sphere geometry, much larger than previous QMC simulations~\cite{wangPhases2021,chenPhases2023,huffman2026generalizingdeconfinedcriticality3d}.

The measured SO(5) order parameter and its corresponding scaling dimension for both spherical and toric geometries are presented in Fig.~\ref{fig:SO5line}. Across both regularization schemes, we identify three distinct regimes as a function of the interaction strength $u$, each mapping to the corresponding regions along the horizontal axis of the phase diagram. At small values of $u < 0.2$, corresponding to the red shaded area in Fig.~\ref{fig:SO5line}, the system resides in the SO(5) symmetry-broken phase. In this regime, the SO(5) order parameter increases with system size, as evidenced by the upward bending of the curves in the log-log plots of panels (a) and (c), which signals spontaneous breaking of SO(5) symmetry in the thermodynamic limit. This behavior is further supported by the scaling dimension, which decreases monotonically with increasing system size in panels (b) and (d), consistent with the expected behavior of a symmetry-broken phase with a finite order parameter in the thermodynamic limit. At intermediate values of $u \sim 0.2$, marked by the orange shaded region in Fig.~\ref{fig:SO5line}, we observe the critical point separating the symmetry-broken and disordered phases. Here, the SO(5) order parameter exhibits clean power-law scaling with system size, accompanied by a scaling dimension that remains constant as a function of inverse system size, the hallmark of a scale-invariant critical state. Interestingly, the two geometries share a similar position of the multicritical point, $\sim 0.2$, as denoted by the orange dot in the universal phase diagram in Fig.~\ref{fig:phasediagram} (d). 

For large values of $u > 0.2$, corresponding to the grey shaded area in Fig.~\ref{fig:SO5line}, the system enters the SO(5)-symmetric disordered phase. In this regime, the SO(5) order parameter retains power-law scaling within accessible system sizes, but with a scaling dimension distinct from that of the multicritical point.
This behavior persists across the full range of $u$ explored; however, up to $u=10,20$ in both geometries, we begin to observe divergent behavior in the scaling dimension. These observations on the one hand confirm that the SO(5) symmetric disordered phase is a stable, extended thermodynamic phase, rather than a finite-size effect or a narrow critical window, on the other hand, suggesting a possible small-gapped phase that only appears up to the sizes we can access (more below). For example, for the system sizes $N_\phi > 50$, the scaling dimension $\Delta_\varv$ increases with increasing system size $N_\phi$ for $u > 0.2$ (black region), while it decreases with increasing $N_\phi$ for $u < 0.2$ (red region), see Figs.~\ref{fig:SO5line}(b) and \ref{fig:SO5line}(d). The sign of the slope of the measured value of $\Delta_\varv$ as a function of $1/N_\phi$, however, indicates the direction of the renormalization group flow, with a positive (negative) slope corresponding to a decreasing (increasing) coupling $u$ towards the infrared. Our observation of a positive slope for $u<0.2$ (red region) together with the negative slope for $u>0.2$ (black region) is therefore inconsistent with the fixed-point annihilation Scenario~2 outlined in Fig.~\ref{fig:phasediagram}(b).

The observed behavior demonstrates the existence of a critical fixed point `M' that separates the ordered and disordered phases.
To extract the scaling dimension $\Delta_{\varv}$ of the SO(5) vector order parameter characterizing the continuous quantum phase transition along the SO(5) symmetric line, we perform linear fits of $\ln(m^2_{\text{SO}(5)})$ versus $\ln(N_\phi)$ for different $u$ values, where the slope of each fit directly yields the corresponding $\Delta_{\varv}$ as dictated by finite-size scaling theory. By progressively excluding small system sizes from the fits, we find that $\Delta_\varv$ decreases rapidly for {$u \ll 0.2$}, a clear signature that the system resides in the SO(5) symmetry-broken phase. In contrast, for {$u \sim 0.2$}, $\Delta_{\varv}$ converges to a finite non-zero value, demonstrating that $m^2_{\text{SO}(5)}$ exhibits clean power-law scaling with $N_\phi$, consistent with the system being in the critical regime. We focus on the parameter points $u=0.175$ (spherical geometry) and $u=0.2$ (toric geometry), where the extracted $\Delta_{\varv}$ shows a flat dependence on $1/N_\phi$, indicating negligible finite-size corrections to the thermodynamic scaling limit. For the largest $N_\phi^{\text{min}}$ (the minimum system size retained in the scaling fits), we obtain $\Delta_{\varv} = 0.52(1)$ and $0.50(3)$ for the spherical and toric geometries, respectively.

We now discuss the disordered phase. 
At the maximal system size available, we do not observe a scaling behavior of the order parameter, as can be seen in Figs.~\ref{fig:SO5line}(b) and \ref{fig:SO5line}(d). 
We observe that $\Delta_{\varv}$ increases with system size. 
At larger $N_{\phi}$, we expect $\Delta_{\varv}$ to stabilize to a larger fixed value.
On a discrete lattice, for a gapped phase, we have $\langle m^2_{\rm SO(5)}\rangle\sim \frac{\xi^d}{R^{2}}$, where $\xi$ is the correlation length and $R$ is the lattice size. 
When using lowest Landau level projection, the number of orbits, $N_{\phi}$ plays the role of an volume $R^2$.
If the disordered phase is a gapped phase, we should get $\langle m^2_{\rm SO(5)}\rangle\sim N_{\phi}^{-1}$ at larger $N_{\phi}$. 
A possible gapped phase consistent with the t'Hooft anomaly is the chiral quantum spin liquid phase, as proposed in~\cite{wangDeconfined2017}.
If the disordered phase is a critical phase, on the other hand, we expect to observe $\langle m^2_{\rm SO(5)}\rangle\sim N_{\phi}^{-\Delta_{\phi}}$.
Even though the non-trivial t'Hooft anomaly forbids a trivial gapped phase, it does not forbid a critical phase, that is, a CFT.  
This phase will be stable.
It has been established that, for such a phase to exist, $\Delta_\varv>0.79$~\footnote{As far as the authors are aware, this bound was independently obtained by David Simmons-Duffin, Yu Nakayama, and Ning Su, respectively. See Section 5.E.4 of~\cite{polandConformal2019} for a detailed discussion.
See also Ref.~\cite{li22} for a recent improved bound. 
}.
Interestingly, reference~\cite{Shackleton:2021fdh} provides a candidate for such a CFT, which is the $Z_2$ gapless spin liquid phase. 
The low-energy effective action for this phase is simply four complex fermions coupled to a $Z_2$ gauge field. 
Determining the precise nature of the SO(5)-disordered phase is an important task that we leave for future work.
We note, however, that the existence of a tricritical point is strongly supported by our numerical result.
The tricritical point discovered here has also been argued to exist in the generalized $J$-$Q$ model~\cite{takahashi2024so5}, even though it was previously inaccessible to quantum Monte Carlo simulations due to the sign problem.
The cross-check between the two different models is also strong evidence.

\begin{figure}[tb!]
\includegraphics[width=\columnwidth]{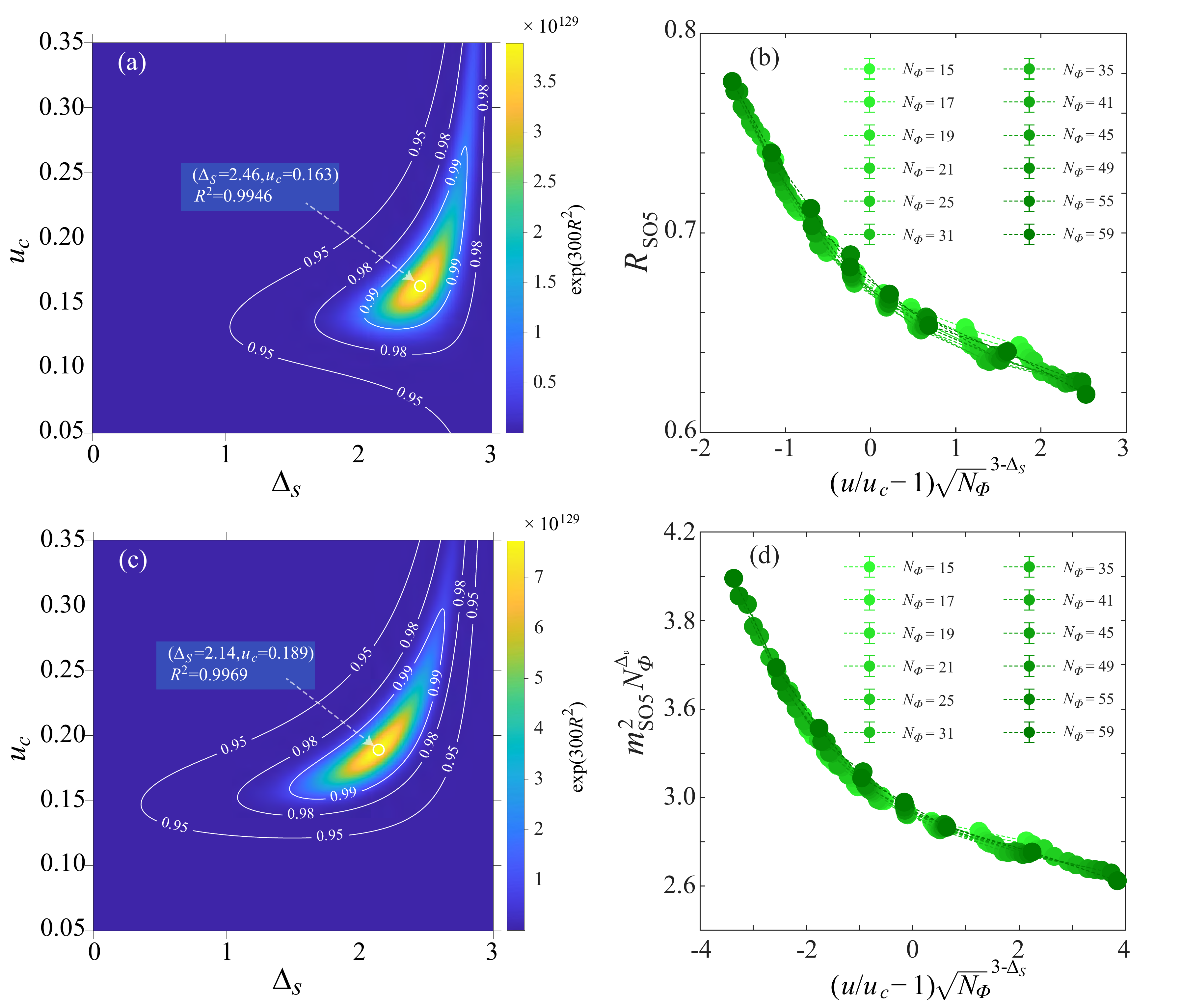}
\caption{\textbf{Finite-size scaling analysis along the SO(5)-symmetric line on the sphere geometry.} (a) $R^2$ goodness-of-fit map from grid scanning of the correlation ratio $R_{\text{SO(5)}}$, as a function of critical coupling $u_c$ and scaling dimension $\Delta_s$. (b) Data collapse of $R_{\text{SO(5)}}$ for $N_\phi \ge 15$, obtained with optimal parameters $u_c=0.163$, $\Delta_s=2.46$. (c) $R^2$ goodness-of-fit map from grid scanning of the squared SO(5) order parameter $m_{\text{SO(5)}}^2$ at fixed $\Delta_\varv=0.53$. (d) Data collapse of $m_{\text{SO(5)}}^2$ using the optimal parameters $u_c=0.189$, $\Delta_s=2.14$ extracted from (c).}
\label{fig:finitesize-scaling-sphere}
\end{figure}

With such understanding, along the SO(5)-symmetric line, we perform a systematic finite-size scaling analysis to precisely locate the quantum critical point separating the SO(5)-broken ordered phase from the SO(5)-symmetric disordered phase, and to estimate the associated critical exponents. Our approach combines $R^2$-optimized grid scanning, data collapse, and crossing-point analysis to robustly extract universal critical properties from finite-size numerical data.

To pinpoint the critical point, we employ the correlation ratio
\begin{equation}
R = 1 - \frac{S(\mathbf{Q} + \Delta \mathbf{q})}{S(\mathbf{Q})},
\end{equation}
where $\mathbf{Q}$ is the ordering wave vector ($\mathbf{Q}=0$ for SO(5), AFM, and VBS order in our regularized geometries), and $\Delta \mathbf{q}$ is the smallest non-zero angular momentum. This correlation ratio is a renormalization-group invariant observable: it approaches unity in the ordered phase and vanishes in the disordered phase, such that crossing points between curves for different system sizes mark the critical point.

We first carry out an $R^2$-optimized grid scanning analysis, as shown in Fig.~\ref{fig:finitesize-scaling-sphere}(a).
Within finite-size scaling theory, correlation ratio $R$ in the vicinity of a quantum critical point obeys the scaling ansatz:
\begin{equation}\label{eq:finite-size-scaling-R}
    R = \mathcal{F}\left( (u/u_{c}-1) \sqrt{N_\phi}^{3-\Delta_s} \right),
\end{equation}
where $\mathcal{F}(x)$ is a universal scaling function and $\Delta_s$ is the scaling dimension of the SO(5) singlet operator~\cite{chenPhases2023}.
We restrict our scaling analysis to the critical regime by including system sizes $N_\phi \ge 15$, with a maximum size of $N_\phi=59$—nearly a threefold increase over the $N_\phi=16$ reached in a prior DMRG study~\cite{chenPhases2023} and the $N_\phi=21$ from an earlier QMC work~\cite{huffman2026generalizingdeconfinedcriticality3d}—and couplings in the range $u \in [0.075, 0.30]$. We approximate $\mathcal{F}(x)$ as a third-order polynomial in the neighborhood of $u_c$, and perform a grid scan over $\Delta_s$ and $u_c$ to identify the parameter set that yields the optimal data collapse, quantified by the coefficient of determination $R^2$. As shown in Fig.~\ref{fig:finitesize-scaling-sphere}(a), the optimal fit gives $u_{c}=0.163$ and $\Delta_s=2.46$.

Using these optimized parameters, we present the resulting data collapse for $R_{\text{SO(5)}}$ in Fig.~\ref{fig:finitesize-scaling-sphere}(b). All data points for $N_\phi \ge 15$ across $u \in [0.075, 0.30]$ collapse onto a single universal curve, further corroborating the critical nature of the transition and validating the extracted critical exponents.

Notably, this value of $u_c$ is fully consistent with the orange region in Fig.~\ref{fig:SO5line}(b), confirming the self-consistency of our scaling procedure and supporting the validity of the scaling dimension $\Delta_\varv=0.52(1)$ extracted from the system-size dependence of the order parameter near the critical point [Fig.~\ref{fig:SO5line}(b)].

A second scaling ansatz, related to the order parameter scaling dimension $\Delta_\varv$, reads
\begin{equation}\label{eq:finite-size-scaling-m}
    m_{\text{SO(5)}}^2 N_\phi^{\Delta_\varv} = \mathcal{F}\left( (u/u_{c}-1) \sqrt{N_\phi}^{3-\Delta_s} \right).
\end{equation}
As shown in Fig.~\ref{fig:finitesize-scaling-sphere}(c), fixing $\Delta_\varv=0.53$ yields an optimal fit with $u_{c}=0.189$ and $\Delta_s=2.14$.
The corresponding data collapse, obtained with these optimized parameters, is displayed in Fig.~\ref{fig:finitesize-scaling-sphere}(d), where all data points again fall onto a single universal curve.

The values of $\Delta_s$ and $u_c$ obtained from the $R^2$-optimized grid scan of $m_{\text{SO(5)}}^2$ are fully consistent with those from the independent analysis of $R_{\text{SO(5)}}$, even when restricting to fits with $R^2>0.99$ as seen in Figs.~\ref{fig:finitesize-scaling-sphere}(a) and \ref{fig:finitesize-scaling-sphere}(c).
Taken together, the results in Figs.~\ref{fig:SO5line} and \ref{fig:finitesize-scaling-sphere} establish a robust, self-consistent picture of a continuous quantum phase transition between the SO(5)-ordered and SO(5)-disordered phases at $u_{c}=0.20(5)$ with $\Delta_s=2.2(6)$.

We have also performed an analogous finite-size scaling analysis on the torus geometry, together with a complementary crossing-point analysis of the SO(5) order parameter correlation ratio for both geometries. These additional results are presented in Supplementary Note~4~\cite{suppl}, and fully confirm the phase structure and critical exponents reported in the main text.

\begin{figure}[tb!]
\includegraphics[width=\columnwidth]{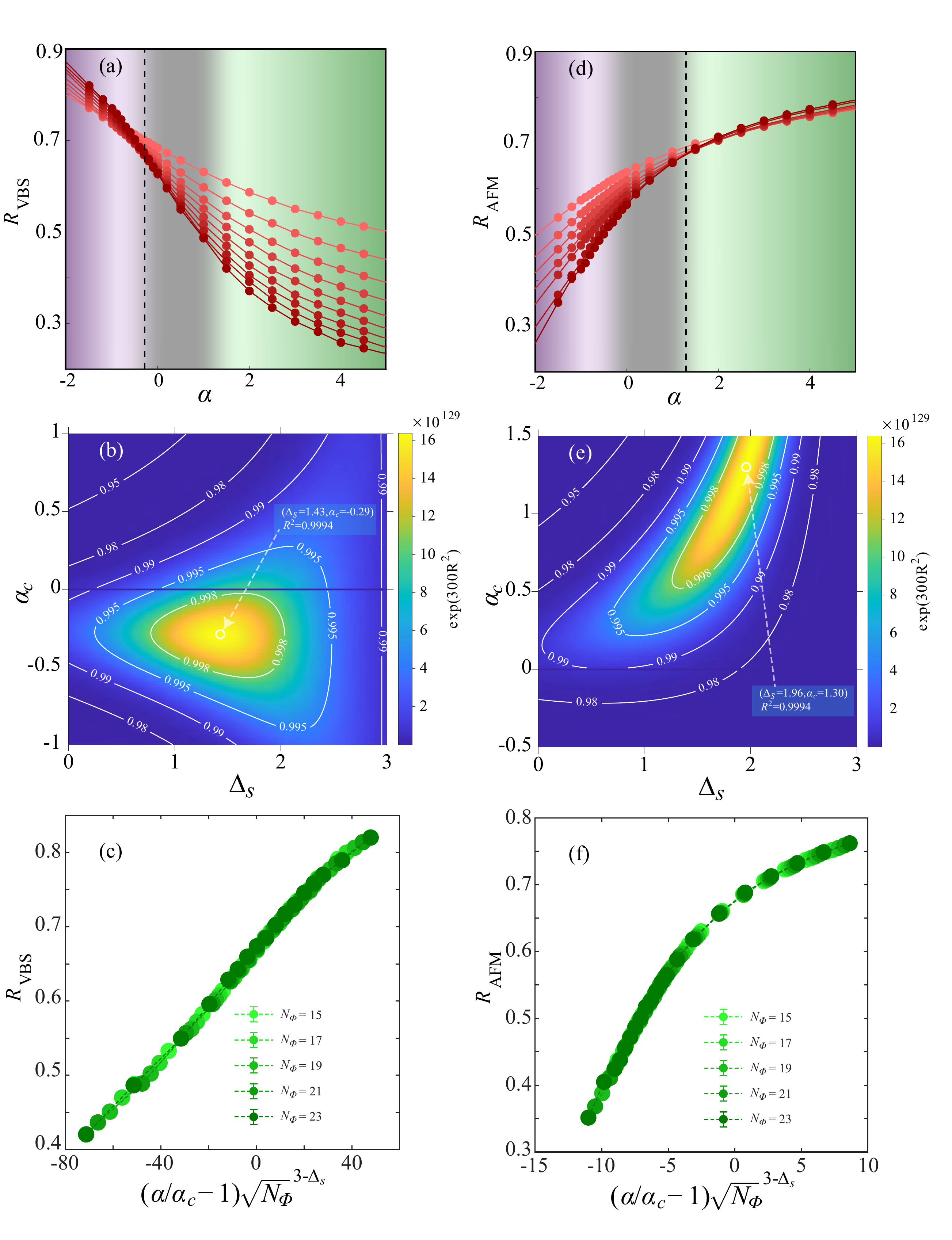}
\caption{\textbf{Phase boundaries and finite-size scaling analysis away from the SO(5)-symmetric line on the sphere geometry at fixed $u=3$.} (a) VBS correlation ratio $R_{\text{VBS}}$ as a function of anisotropy $\alpha$ for increasing system sizes, with cyan, gray, and green backgrounds marking the VBS, disordered, and AFM phases, respectively. Curve crossings identify the VBS–disorder transition. The vertical dash line represents $\alpha_{c,\text{VBS}} = -0.29$. (b) $R^2$ goodness-of-fit map from grid scanning of $R_{\text{VBS}}$. (c) Data collapse of $R_{\text{VBS}}$ obtained with the optimal critical parameters. (d)–(f) Corresponding analysis for the disorder–AFM transition: (d) AFM correlation ratio versus $\alpha$, (e) $R^2$ goodness-of-fit map, and (f) data collapse of the AFM correlation ratio. The vertical dash line in (d) represents $\alpha_{c,\text{AFM}} = 1.3$. 
Panels (a) and (d) display data for system sizes $N_\phi = 7, 9, 11, 15, 17, 19, 21, 23$, with curves colored from light to dark as $N_\phi$ increases.
}
\label{fig:awaySO5_sphere}
\end{figure}

\subsection{Phase boundaries away from the SO(5)-symmetric line}
\label{sec:results_phaseboundaries}
Finally, we determine the phase boundaries away from the SO(5)-symmetric line, which set the vertical extent of the disordered phase shown in Fig.~\ref{fig:phasediagram}. We focus on the sphere geometry at fixed $u=3$, deep within the SO(5)-symmetric disordered phase, and tune the anisotropy $\alpha$ to map the transitions into the VBS and AFM ordered phases. Our results for the VBS–disorder and disorder–AFM transitions are summarized in Fig.~\ref{fig:awaySO5_sphere}.

In Fig.~\ref{fig:awaySO5_sphere}(a), the cyan, gray, and green regions denote the VBS, disordered, and AFM phases, respectively. The VBS correlation ratio $R_{\text{VBS}}$ exhibits clear crossing points across system sizes at a finite negative $\alpha$, demonstrating that the VBS–disorder transition occurs at a finite offset from the SO(5)-symmetric line. We then perform an $R^2$-optimized grid scanning analysis, following the same protocol used along the SO(5) line, with results presented in Fig.~\ref{fig:awaySO5_sphere}(b). We restrict the scaling analysis to the critical regime using system sizes $N_\phi \ge 15$, and approximate the universal scaling function $\mathcal{F}(x)$ by a third-order polynomial. The grid scan over the critical anisotropy $\alpha_{c,\text{VBS}}$ and scaling dimension $\Delta_{S,\text{VBS}}$, quantified by the coefficient of determination $R^2$, yields best-fit values $\alpha_{c,\text{VBS}} = -0.29$ and $\Delta_{S,\text{VBS}} = 1.43$. Using these optimized parameters, the resulting data collapse for $R_{\text{VBS}}$ is shown in Fig.~\ref{fig:awaySO5_sphere}(c). All data points fall cleanly onto a single universal curve, consistent with a continuous quantum phase transition between the VBS phase and the SO(5)-symmetric disordered phase.

Analogously, results for the disorder–AFM transition are presented in Figs.~\ref{fig:awaySO5_sphere}(d)–(f). The same combined analysis of crossing points, $R^2$-optimized grid scanning, and data collapse yields best-fit parameters for the disorder–AFM transition: $\alpha_{c,\text{AFM}} = 1.3$ and $\Delta_{S,\text{AFM}} = 1.96$.

Taken together, our analyses of the VBS–disorder and disorder–AFM transitions strongly support the existence of an extended disordered phase emanating from the critical point on the SO(5)-symmetric line, which remains stable against weak anisotropy perturbations parametrized by $\alpha$. Consistent results on the torus geometry are presented in Supplementary Note~1~\cite{suppl}.

We note, however, that the magnitudes of both $\alpha_{c,\text{VBS}}$ and $\alpha_{c,\text{AFM}}$ are small, and their $3\sigma$ error bars approach or even encompass zero. On the basis of our current numerical data, we therefore cannot rule out the possibility that the disordered phase is only weakly stable away from the SO(5) line, or that it may be confined strictly to the SO(5)-symmetric line. For the same reason, in Fig.~\ref{fig:phasediagram}(d) we only show the best-fit locations of these transition points as red diamonds without explicit error bars.
If the disordered phase exists in an extended region that contains the SO(5)-invariant line, one natural candidate is the chiral quantum spin liquid~\cite{wangDeconfined2017}, another is the so-called $Z_2$ gapless spin liquid phase as proposed in~\cite{Shackleton:2021fdh}.
If the disordered phase is instead confined strictly to the SO(5)-symmetric line, this leads to an interesting scenario. 
As mentioned in the previous section, the t'Hooft anomaly does not rule out the possibility of an SO(5)-invariant conformal field theory. 
It is precisely because of this that the phase transition of the original $J$-$Q$ model was long believed to be second order.
This was only ruled out later because the measured critical exponents from large-size quantum Monte Carlo simulations~\cite{Sandvik:2020yup} are below the conformal bootstrap bounds~\cite{nakayamaNecessary2016}. 
These studies still leave space for a stable SO(5)- invariant conformal field theory, with large enough critical exponents satisfying the bootstrap bounds, that can flow to N\'eel or VBS phase upon turning on the SO(5) breaking interactions. 
This CFT will be a candidate for the disordered phase if it is confined strictly to the SO(5)-symmetric line. 
It corresponds to the fixed point D in Scenario 1, as shown in Fig.~\ref{fig:phasediagram}(a).

\section{Discussion}
\label{sec:discussion}
In this work, we developed a continuous field QMC with Langevin global update, to explore the phase diagram of the projected Landau level of half-filled Dirac fermions with SO(5) symmetry on spherical and toric geometries. 
This model provides a faithful lattice regularization of the nonlinear sigma model with a Wess-Zumino-Witten term in 2+1 dimensions~\cite{maTheory2020, nahumNote2020, zou21, hawashin25}, and captures the physics of the deconfined quantum phase transition~\cite{senthilQuantum2004, senthilDeconfined2023}.
By reducing the computational complexity by at least an order of $O(N_\phi)$ on torus and sphere geometries and by purposely exploring the phase space away from the SO(5) symmetric line, we found the universal structure of the phase diagram (as shown in Fig.~\ref{fig:phasediagram} (d)) on both geometries, in that, besides the SO(5) symmetry-breaking phase and the multi-critical point from which the N\'eel and VBS phases originate, there exists a stable and SO(5) symmetric disorder phase (either gapped or gapless) that separates the N\'eel and VBS phases.

In contrast to previous QMC studies of quantum spin systems~\cite{louvbsneel2009,kaulLattice2012,blockFate2013,pujari13,pujari15} and classical loop models~\cite{nahumDeconfined2015,nahumEmergent2015}, our regularization provides direct access to the SO(5)-symmetric subspace without fine-tuning. 
This enables a complete characterization of the phase structure, comprising an ordered phase at small coupling $u$, a disordered phase at large $u$, and an intervening critical point. Upon explicitly breaking SO(5) symmetry down to $\mathrm{SU}(2)\times\mathrm{U}(1)$, this critical point becomes multicritical.
Within the disordered phase, we observe drifting and enhancement of the scaling dimension similar to those reported in earlier studies of spin and loop models~\cite{louvbsneel2009,kaulLattice2012,blockFate2013,pujari13,pujari15,nahumDeconfined2015,nahumEmergent2015}. This indicates that the correlation length of the gapped modes is comparable with the system size, covering the true nature of this disordered phase.
By contrast, we do not observe comparable drift at the multicritical point, indicating that it is not affected by a putative fixed-point annihilation mechanism, in contrast to earlier expectations~\cite{maTheory2020,nahumNote2020, zou21, hawashin25, ray25, zhouThe2024,huffman2026generalizingdeconfinedcriticality3d,Feuerpfeil2026,Maity2026}. 
At this point, we find {$\Delta_\varv = 0.52(1)$ and $0.50(3)$} for the spherical and toric geometries, respectively.
Most importantly, the multicritical point separates a small-coupling regime in which the measured scaling dimension $\Delta_\varv$ decreases with system size from a strong-coupling regime in which $\Delta_\varv$ increases with system size for sufficiently large $N_\phi$. This refutes the fixed-point annihilation Scenario~2 outlined in Fig.~\ref{fig:phasediagram}(b).

The scaling dimensions for the SO(5) vector $\Delta_\varv = 0.52(1)$ and the SO(5) singlet $\Delta_s = 3 - 1/\nu = 2.2(6)$ provide strong numerical evidence for a single interacting fixed point `M' at finite coupling and a slow renormalization group flow in the strong-coupling regime.
In particular, we emphasize that the measured values for both $\Delta_\varv$ and $\Delta_s$ are consistent with conformal bootstrap bounds~\cite{polandConformal2019,li22}, as the SO(5) singlet remains a relevant operator ($\Delta_s < 3$), but $\Delta_s > 1.044$ (equivalent to $\nu > 0.511$). 
To our knowledge, this constitutes the first direct large-scale numerical determination of both $\Delta_\varv = (1+\eta_\varv)/2$ and $\Delta_s = 3 - 1/\nu$ right at the multicritical point.
Recent conformal bootstrap studies~\cite{chesterBootstrapping2024} further show that a precise value of $\Delta_\varv$ can be used to constrain the scaling dimensions of additional operators, including the SO(5) singlet $s$, the rank-2 tensor $t$, and the rank-3 tensor $t_3$. While existing analyses rely on the large-$N$ estimate $\Delta_\varv \approx 0.630$~\cite{dyer15}, our results suggest that the true value at the multicritical point is smaller. It would therefore be particularly interesting to revisit these bootstrap predictions for $\Delta_s$, $\Delta_t$, and $\Delta_{t_3}$ using the updated estimate.

Our phase diagram also bears a striking similarity with that of the generalized $J$-$Q$ model~\cite{takahashi2024so5}. In particular, a first-order transition across the SO(5) breaking phase, together with a multicritical point and a disordered phase, separates the N\'eel and VBS phases.

The universal phase diagram that we observe is not only of theoretical importance but also resonates with the recently findings in the Shastry-Sutherland Mott insulator SrCu$_2$(BO$_3$)$_2$, where the existence of a symmetric intermediate phase (as our disorder phase) at low temperature, occupying a large parameter space between the plaquette-singlet (VBS) and antiferromagnetic (N\'eel) phases, has been found both by the thermal tensor-network simulations on Shastry-Sutherland model at finite magnetic field~\cite{songThermodynamics2026} and by tensor-network and neural quantum state at zero field (with narrower range)~\cite{yangQuantum2022,viterittiTransformer2025,maity2025evidence,corbozQuantum2025}, as well as the high-pressure calorimetry measurements under magnetic field~\cite{Guo2020QuantumPhases,Cui2023Proximate,Guo2025Deconfined,Zayed20174-spin,Jiménez2021analogue,guoT-linear2026}. 

It also resonates with the deconfined phase in the setting of emergent U(1) gauge field coupled with spinons, such that a stable U(1) Dirac spin liquid phase could exist both from lattice model simulations~\cite{xuMonte2019,chenEmergent2025,fengScalable2026} (up to the space-time scale of $60\times60\times600$ with state-of-the-art hybrid Monte Carlo implementation on GPU) and from triangular lattice antiferromagnets
YbZn$_2$GaO$_5$~\cite{BagEvidence24,WuSpinDynamics25} and the $A$-YbSe$_2$
delafossites~\cite{ScheieKYbSe2, ScheieNaYbSe2} as well as the kagom\'e
antiferromagnet
YCu$_3$(OD)$_6$Br$_2$[Br$_{x}$(OD)$_{1-x}$]~\cite{zengSpectral2024,hanSpin2024},
where specific heat measurements revealed the possible symmetric (as our disorder) phase of the Mott insulating material~\cite{zengPossible2022}, and inelastic neutron scattering experiments have found spectra that are consistent with a Dirac cone filled with a continuum of excitations. The transition from the (proximate) Dirac spin liquid (possibly with a very small gap beyond experiment resolution) to AFM and VBS phases could further motivate the theoretical understanding of the Scenario 3 and the universal phase diagram in Fig.~\ref{fig:phasediagram} (d), as well as its experimental realization in quantum materials.

The methodology we have developed here, i.e., the global update in the continuous auxiliary field QMC simulation to reduce the computational complexity, will also be useful for the investigations of the interaction effect in fractional quantum Hall systems~\cite{assaadFractional1995,wangHybrid2026} and the ideal Chern band systems~\cite{bandRoy2014,wangExact2021,topologicalValentin2025} which are good approximations to experimentally realized quantum moiré materials such as twisted bilayer graphene~\cite{Cao2018insulator} and MoTe$_2$~\cite{caiSignatur2023} and other fractional quantum (anomalous) Hall systems~\cite{hanCorrelated2024,luFractional2024}. In these systems, the projected interactions onto different basis functions (Aharonov-Casher bands or Landau levels) naturally introduce long-range interactions that render the conventional local update in QMC inefficient. We note that the first attempts with global update in the continuous auxiliary field in this direction have been taken~\cite{angleHuang2024,wangHybrid2026,huangStrain2026,shanTrion2026}, and more promising results are expected.

\section{Methods}
\label{sec:methods}

\subsection{Continuum SO(5) model and regularization schemes}
\label{sec:methods_continuum}
\subsubsection{Continuum Hamiltonian}
Let's consider a continuum Hamiltonian for interacting Dirac fermions in a magnetic field, projected onto the LLL. After integrating out high-energy fermionic degrees of freedom, the system is described by a (2+1)D SO(5) nonlinear sigma model with a level-1 Wess-Zumino-Witten term\cite{leeWess2015,ippolitiHalf2018}, with the microscopic Hamiltonian
\begin{equation}
\label{eq:hamiltonian_continuum_methods}
\hat{\mathcal{H}}=  \int d\boldsymbol{r} \left[ {U_0}\left(\hat{\boldsymbol{\psi}}^{\dagger}_{\boldsymbol{r}} \hat{\boldsymbol{\psi}}_{\boldsymbol{r}}-C_{\boldsymbol{r}}\right)^2
 -\sum_{i=1}^5 U_i \left(\hat{\boldsymbol{\psi}}^{\dagger}_{\boldsymbol{r}} \Gamma^i \hat{\boldsymbol{\psi}}_{\boldsymbol{r} }\right)^2 \right],
\end{equation}
where $\hat{\boldsymbol{\psi}}_{\boldsymbol{r}}$ is a four-component Dirac spinor encoding the valley ($\boldsymbol{\tau}$) and spin ($\boldsymbol{\sigma}$) degrees of freedom of the Dirac fermions. 
The term proportional to $U_0$ enforces half-filling via a charge density constraint, with $C_{\boldsymbol{r}}$ a background charge density.

To facilitate numerical simulation, we project the continuum field operator onto the LLL, which discretizes the model in momentum space. The field operator is expanded as
\begin{equation}
\hat{\psi}_{\boldsymbol{r}} = \sum_{k=1}^{N_\phi} \phi_{k}(\boldsymbol{r}) \mathbf{c}_{k},
\end{equation}
where $\phi_{k}(\boldsymbol{r})$ are the normalized LLL single-particle wave functions, $N_\phi$ is the degeneracy of the LLL (equal to the number of magnetic flux quanta piercing the system, i.e., the system size), and $\mathbf{c}_{k}$ are the four-component fermionic annihilation operators in the LLL basis. Substituting this expansion into Eq.~\eqref{eq:hamiltonian_continuum_methods}, we obtain the regularized Hamiltonian in momentum space:
\begin{equation}
\mathcal{H} = U_0 \mathcal{H}_0 -\sum_{i=1}^5 U_i \mathcal{H}_i,
\label{eq:model}
\end{equation}
where
\begin{equation}
\mathcal{H}_i=\sum_{\mathbf{q}}^{N_{\mathbf{q}}} n^{\Gamma^{i}}_{\mathbf{q}} n^{\Gamma^{i}}_{-\mathbf{q}}
\end{equation}
for $i=0,1,\dots,5$. Here, $n^{\Gamma^{i}}_{\mathbf{q}} = \int d\boldsymbol{r} e^{i\mathbf{q}\cdot\boldsymbol{r}} \hat{\boldsymbol{\psi}}^{\dagger}_{\boldsymbol{r}} \Gamma^i \hat{\boldsymbol{\psi}}_{\boldsymbol{r}}$ is the Fourier-transformed density operator for the $i$-th SO(5) component, and $N_{\mathbf{q}}$ is the number of momentum points included in the simulation. The explicit form of $n^{\Gamma^{i}}_{\mathbf{q}}$ depends on the geometry of the system, as detailed below.

\subsubsection{Toric regularization}
For the toric geometry, we consider a 2D system of size $L_x\times L_y$ with periodic boundary conditions, subject to a uniform perpendicular magnetic field in the Landau gauge. The LLL wave functions are plane-wave-like states with a Gaussian envelope, given by
\begin{equation}
    \phi_{k}(\boldsymbol{r})=\frac{1}{\sqrt{L_y l_B \sqrt{\pi}}} e^{-\left(r_x / l_B-l_B 2\pi k / L_y \right)^2 / 2} e^{i 2\pi k r_y/ L_y},
\end{equation}
where $l_B = \sqrt{\phi_0/(2\pi|B|)}$ is the magnetic length (set to 1 in our simulations), $\phi_0=h/e$ is the magnetic flux quantum, and $k=1,2,\dots,N_\phi$ labels the degenerate LLL states, with $N_\phi = |B|L_xL_y/\phi_0$.

The momentum points are parameterized by integer indices $(q_x,q_y)$, with $\mathbf{q} = (q_x \sqrt{2\pi/N_\phi}, q_y \sqrt{2\pi/N_\phi})$. The density operator in this basis takes the form
\begin{widetext}
\begin{equation}
    n^{\Gamma^{i}}_{\mathbf{q}} =\frac{e^{-\frac{2\pi}{4N_\phi}\left(q_x^2+q_y^2\right) }}{4\pi\sqrt{N_\phi}}  \sum_{k=1}^{N_\phi} \sum_{a,b=1}^4 e^{\frac{i}{2}\left(2 k-q_y\right) q_x (2\pi/N_\phi)}  \left( {c}^{\dagger}_{k,a} \Gamma^{i}_{a,b} {c}_{k-q_y,b} - 2 \delta_{q_y,0} \delta_{i,0} \right),
\label{eq:torus_density_operator_methods}
\end{equation}
\end{widetext}
where $\delta_{q_y,0} \delta_{i,0}$ is a background term that enforces particle-hole symmetry at half filling. We only include momenta satisfying $e^{-\frac{2\pi}{4N_\phi}(q_x^2 + q_y^2)} > 0.01$, such that $N_{\mathbf{q}} \propto N_\phi$.

\subsubsection{Spherical regularization}
For the spherical geometry, we consider Dirac fermions on the surface of a sphere pierced by a magnetic monopole of charge $4\pi s$ (where $s$ is an integer or half-integer). The degeneracy of the LLL is $N_\phi=2s+1$, with wave functions given by monopole harmonics:
\begin{equation}
    \phi_k(\boldsymbol{r})=N_{m_k} e^{i m_k \phi} \cos ^{s+m_k}\left(\frac{\theta}{2}\right) \sin ^{s-m_k}\left(\frac{\theta}{2}\right),
\end{equation}
where $(\theta,\phi)$ are the spherical coordinates, $m_k = k-1-s$, and $N_{m_k}=\sqrt{\frac{(2 s+1)!}{4 \pi(s+{m_k})!(s-{m_k})!}}$ is the normalization constant.

On the sphere, momentum is naturally parameterized by angular momentum indices $(l,m)$, with $l\in\{0,1,\dots,2s\}$ and $m\in\{-l,-l+1,\dots,l\}$. The density operator in this basis is expressed in terms of Wigner 3-j symbols as
\begin{widetext}
\begin{equation}
    n^{\Gamma^{i}}_{l,m} = \sum_{k=1}^{N_\phi} \sum_{a,b=1}^4   (-1)^{s+k+m}   \frac{(2s+1)\sqrt{2l+1}}{2}\left(\begin{array}{ccc}
s & l & s \\
-k & -m & k+m
\end{array}\right)\left(\begin{array}{ccc}
s & l & s \\
-s & 0 & s
\end{array}\right) \left( {c}^{\dagger}_{k,a} \Gamma^{i}_{a,b} {c}_{k+m,b} - 2 \delta_{m,0} \delta_{i,0} \right),
\label{eq:sphere_density_operator_methods}
\end{equation}
\end{widetext}
where $\delta_{m,0} \delta_{i,0}$ again enforces particle-hole symmetry. We include all angular momentum channels, giving $N_{\mathbf{q}}=(N_\phi+1)^2$.

\subsection{Interaction decoupling via Fierz identity}
\label{sec:methods_fierz}
The first key step in our algorithm is to transform the original Hamiltonian in Eq.~\eqref{eq:model} into an equivalent form with decoupled spin degrees of freedom, eliminating the fermionic sign problem and simplifying the interaction structure. This is achieved via the Fierz identity for SU(4) generators, which form a complete basis for $4\times4$ matrices (describing the 2 valley × 2 spin degrees of freedom of our Dirac fermions).

\begin{table*}[htp!]
\centering
\caption{The commutation relations between the SU(4) generator matrices.}
\label{tab:wide_table} 
\begin{tabular}{|*{17}{c|}} 
\hline
\  & $T^1$ & $T^2$ & $T^3$ & $T^4$ & $T^5$ & $T^6$ & $T^7$ & $T^8$ & $T^9$ & $T^{10}$ & $T^{11}$ & $T^{12}$ & $T^{13}$ & $T^{14}$ & $T^{15}$ & $T^{16}$ \\ \hline
$T^{13}$ & +1 & +1 & +1 & +1 & -1 & -1 & -1 & -1 & -1 & -1 & -1 & -1 & +1 & +1 & +1 & +1 \\ \hline
$T^{14}$ & +1 & +1 & -1 & -1 & -1 & -1 & +1 & +1 & -1 & -1 & +1 & +1 & +1 & +1 & -1 & -1 \\ \hline
$T^{15}$ & +1 & -1 & +1 & -1 & -1 & +1 & -1 & +1 & -1 & +1 & -1 & +1 & +1 & -1 & +1 & -1 \\ \hline
$T^{16}$ & +1 & -1 & -1 & +1 & -1 & +1 & +1 & -1 & -1 & +1 & +1 & -1 & +1 & -1 & -1 & +1 \\ \hline
\end{tabular}
\end{table*}

The basic concept of Fierz identiy is that any $4\times 4$ matrix can be expanded by 16 matrices. These 16 matrices is the SU(4) generators defined as
\begin{equation}
\begin{aligned}
       T^{i} \in \{ & I_2\bigotimes I_2, I_2\bigotimes \sigma_x, I_2\bigotimes \sigma_y, I_2\bigotimes \sigma_z, \\
       & \tau_x \bigotimes I_2, \tau_x\bigotimes \sigma_x, \tau_x \bigotimes \sigma_y, \tau_x\bigotimes \sigma_z, \\
       & \tau_y\bigotimes I_2, \tau_y\bigotimes \sigma_x, \tau_y\bigotimes \sigma_y, \tau_y\bigotimes \sigma_z, \\
       & \tau_z\bigotimes I_2, \tau_z\bigotimes \sigma_x, \tau_z\bigotimes \sigma_y, \tau_z\bigotimes \sigma_z  \},
\end{aligned}
\end{equation}
where $i$ runs from 1 to 16.
The SU(4) generators $\left\{T^i\right\}=\left\{\sigma^a \otimes \tau^b\right\}$ (with $a, b \in\{0,1,2,3\}$) form a complete basis for $4\times 4$ matrices. Any arbitrary $4\times 4$ matrix $A$ can thus be expanded as
\begin{equation}
A_{\alpha \beta}=\sum_i c_i T_{\alpha \beta}^i
\end{equation}
where $c_i=\frac{1}{4} \operatorname{Tr}\left(T^i A\right)$ follows from the orthogonality condition $\operatorname{Tr}\left(T^i T^j\right)=4 \delta_{i j}$.
Consider the tensor product $T^i \otimes T^i$ as a $16\times16$ matrix. Expanding it in the basis $\left\{T^j \otimes T^k\right\}$, we write 
\begin{equation}\label{eq:fi1}
T_{\alpha \beta}^i T_{\gamma \eta}^i=\sum_{j, k} b_{i j k} T_{\alpha \eta}^j T_{\gamma \beta}^k,
\end{equation}
where the indices $\beta$ and $\eta$ are exchanged to reorganize the matrix structure.
To know $b_{ijk}$, we could multiply $T_{\eta,\alpha}^m$ and $T_{\beta,\gamma}^n$ on both side of Eq.~\eqref{eq:fi1} then take the trace over all indices, obtaining
\begin{equation}
\sum_{\alpha \beta \gamma \eta}  T_{\eta \alpha}^m T_{\alpha \beta}^i T_{\beta \gamma}^n T_{\gamma \eta}^i=\sum_{j, k} b_{i j k}\left(\sum_{\alpha \eta} T_{\alpha \eta}^j T_{\eta \alpha}^m\right)\left(\sum_{\beta \gamma} T_{\gamma \beta}^k T_{\beta \gamma}^n\right).
\label{eq:eq13}
\end{equation}
The LHS of Eq.~\eqref{eq:eq13} simplifies to $\operatorname{Tr}\left(T^m T^i T^n T^i \right)$, while the orthogonality condition $\operatorname{Tr}\left(T^j T^m\right) = 4\delta_{jm}$ and $\operatorname{Tr}\left(T^k T^n\right) = 4\delta_{kn}$ reduces the RHS to $16 \ b_{imn}$. For SU(4) generators, $T^i$ either commutes or anticommutes with $T^n$
\begin{equation}
    T^i T^n= \pm T^n T^i \Longrightarrow T^i T^n T^i= \pm T^n.
\end{equation}
Substituting this into the trace, we find
\begin{equation}
\operatorname{Tr}\left(T^m T^i T^n T^i \right)= \pm \operatorname{Tr}\left(T^m T^n\right)= \pm 4 \delta_{m n} .
\end{equation}
Thus, the coefficients become $b_{i m n}=\pm  \frac{ 1}{4} \delta_{m n}$.
Substituting $b_{imn}$ back into Eq.~\eqref{eq:fi1}, we obtain
\begin{equation}\label{eq:fi2}
T_{\alpha \beta}^i T_{\gamma \eta}^i=\sum_{j} \pm \frac{1}{4} T_{\alpha \eta}^j T_{\gamma \beta}^j,
\end{equation}
where the $\pm$ sign depends on the commutation properties of $T^i$ and $T^j$.

Next, we could insert two 4 components fermionic vector $\mathbf{c} = \{c_1,c_2,c_3,c_4\} $ and two $\mathbf{c}^\dagger$ in Eq.~\eqref{eq:fi2} at proper position and, after taking the trace over all indices, obtain
\begin{equation}\label{eq:fi3}
\begin{aligned}
    \left(\mathbf{c}^\dagger T^i \mathbf{c}\right)^2& = \sum_{\alpha \beta \gamma \eta} c_\alpha^\dagger T_{\alpha \beta}^i c_\beta  c_\gamma^\dagger  T_{\gamma \eta}^i c_\eta \\
    & = \sum_{\alpha \beta \gamma \eta} \sum_{j} \pm\frac{1}{4}  T_{\alpha \eta}^j T_{\gamma \beta}^j c_\alpha^\dagger  c_\beta  c_\gamma^\dagger  c_\eta.
\end{aligned}
\end{equation}
By the fermionic anticommutation relations, we have 
\begin{equation}
    c_\alpha^\dagger  c_\beta  c_\gamma^\dagger  c_\eta = c_\alpha^\dagger  c_\beta \delta_{\gamma \eta} + c_\alpha^\dagger  c_\eta \delta_{\beta \gamma} - c_\alpha^\dagger  c_\eta  c_\gamma^\dagger  c_\beta,
\end{equation}
then
\begin{equation}
    \begin{aligned}
        & \sum_{\alpha \beta \gamma \eta} T_{\alpha \eta}^j T_{\gamma \beta}^j c_\alpha^\dagger  c_\beta  c_\gamma^\dagger  c_\eta =  \sum_{\alpha \beta \gamma \eta} T_{\alpha \eta}^j T_{\gamma \beta}^j c_\alpha^\dagger  c_\beta \delta_{\gamma \eta} \\
        & + \sum_{\alpha \beta \gamma \eta} T_{\alpha \eta}^j T_{\gamma \beta}^j c_\alpha^\dagger  c_\eta \delta_{\beta \gamma} - \sum_{\alpha \beta \gamma \eta} T_{\alpha \eta}^j T_{\gamma \beta}^j 
        c_\alpha^\dagger  c_\eta  c_\gamma^\dagger  c_\beta \\
        &=  \sum_{\alpha \beta \gamma} T_{\alpha \gamma}^j T_{\gamma \beta}^j c_\alpha^\dagger  c_\beta + \sum_{\alpha \beta \eta}  T_{\alpha \eta}^j T_{\beta \beta}^j c_\alpha^\dagger  c_\eta - \left(\mathbf{c}^\dagger T^j \mathbf{c}\right)^2.
        \end{aligned}
\end{equation}

Based on these derivation and $\sum_{\alpha \beta \gamma} T_{\alpha \gamma}^j T_{\gamma \beta}^j c_\alpha^\dagger  c_\beta  = \sum_{\alpha \beta} c_\alpha^\dagger I_{\alpha\beta}   c_\beta  =  \left(\mathbf{c}^\dagger I_4 \mathbf{c}\right)$ because of $T^j T^j = I_4$ for every $j$; we know $\sum_{\alpha \beta \eta}  T_{\alpha \eta}^j T_{\beta \beta}^j c_\alpha^\dagger  c_\eta = \left(\mathbf{c}^\dagger T^j \mathbf{c}\right) \delta_{j1} $ because of $\operatorname{Tr}(T^j) = 4\delta_{j1}$.

Then we have 
\begin{equation}\label{eq:fi4}
    \sum_{\alpha \beta \gamma \eta} T_{\alpha \eta}^j T_{\gamma \beta}^j c_\alpha^\dagger  c_\beta  c_\gamma^\dagger  c_\eta = \left(\mathbf{c}^\dagger I_4 \mathbf{c}\right) + \left(\mathbf{c}^\dagger T^j \mathbf{c}\right) \delta_{j1} - \left(\mathbf{c}^\dagger T^j \mathbf{c}\right)^2.
\end{equation}    
Substituting Eq.~\eqref{eq:fi4} back into Eq.~\eqref{eq:fi3}, we obtain
\begin{equation}\label{eq:fi5}
     \left(\mathbf{c}^\dagger T^i \mathbf{c}\right)^2 =  \sum_j \pm\frac{1}{4} \left[ \left(\mathbf{c}^\dagger I_4 \mathbf{c}\right) + 4\left(\mathbf{c}^\dagger T^j \mathbf{c}\right) \delta_{j1} - \left(\mathbf{c}^\dagger T^j \mathbf{c}\right)^2 \right].
\end{equation}

Our Hamiltonian $\mathcal{H}$, i.e., Eq.~\eqref{eq:model}, includes the four fermionic operators $\left(\mathbf{c}^\dagger \Gamma^i \mathbf{c}\right)^2$ with $i$ running from $0$ to $5$. We have $\Gamma^0 = T^1$, $\Gamma^1 = T^5$, $\Gamma^2 = T^9$, $\Gamma^3 = T^{14}$,  $\Gamma^4 = T^{15}$ and $\Gamma^5 = T^{16}$. As discussed above, we would like to use the combination of $\left(\mathbf{c}^\dagger T^1 \mathbf{c}\right)^2$, $\left(\mathbf{c}^\dagger T^5 \mathbf{c}\right)^2$, $\left(\mathbf{c}^\dagger T^9 \mathbf{c}\right)^2$ and $\left(\mathbf{c}^\dagger T^{13} \mathbf{c}\right)^2$ to substitute $\left(\mathbf{c}^\dagger T^{14} \mathbf{c}\right)^2 + \left(\mathbf{c}^\dagger T^{15} \mathbf{c}\right)^2+ \left(\mathbf{c}^\dagger T^{16} \mathbf{c}\right)^2$.
To understand this substitution, it's useful to list the commutation relation between $T^i$ as given in Tab.~\ref{tab:wide_table}. In Tab.~\ref{tab:wide_table}, $+1$ or $-1$ in the i-th row and j-th column represents $T^i$  commutes or anticommutes with $T^j$, and there is a correspondence with the $\pm1$ in Eq.~\eqref{eq:fi5}.

\begin{widetext}
Then we could obtain that 
    \begin{equation}
    \left(\mathbf{c}^\dagger T^{13} \mathbf{c}\right)^2 + \left(\mathbf{c}^\dagger T^{14} \mathbf{c}\right)^2 + \left(\mathbf{c}^\dagger T^{15} \mathbf{c}\right)^2 + \left(\mathbf{c}^\dagger T^{16} \mathbf{c}\right)^2 = 4 \left(\mathbf{c}^\dagger T^1 \mathbf{c}\right)  - \left[ \left(\mathbf{c}^\dagger T^{1} \mathbf{c}\right)^2 - \left(\mathbf{c}^\dagger T^{5} \mathbf{c}\right)^2 - \left(\mathbf{c}^\dagger T^{9} \mathbf{c}\right)^2 + \left(\mathbf{c}^\dagger T^{13} \mathbf{c}\right)^2 \right].
\end{equation}
Or equivalently 
    \begin{equation}\label{eq:fieq}
    - \left(\mathbf{c}^\dagger T^{14} \mathbf{c}\right)^2 - \left(\mathbf{c}^\dagger T^{15} \mathbf{c}\right)^2 - \left(\mathbf{c}^\dagger T^{16} \mathbf{c}\right)^2 =   \left(\mathbf{c}^\dagger T^{1} \mathbf{c} - 2\right)^2 - \left(\mathbf{c}^\dagger T^{5} \mathbf{c}\right)^2 - \left(\mathbf{c}^\dagger T^{9} \mathbf{c}\right)^2 + 2\left(\mathbf{c}^\dagger T^{13} \mathbf{c}\right)^2 - 4.
\end{equation}

The Hamiltonian $\mathcal{H}$ in Eq.~\eqref{eq:model} can be represented as $ U_0\left(\mathbf{c}^\dagger T^{1} \mathbf{c} - 2\right)^2 - U_K [\left(\mathbf{c}^\dagger T^{5} \mathbf{c}\right)^2 + \left(\mathbf{c}^\dagger T^{9} \mathbf{c}\right)^2] - U_N [\left(\mathbf{c}^\dagger T^{14} \mathbf{c}\right)^2 + \left(\mathbf{c}^\dagger T^{15} \mathbf{c}\right)^2 + \left(\mathbf{c}^\dagger T^{16} \mathbf{c}\right)^2]$. 
Using Eq.~\eqref{eq:fieq} and dropping an overall constant, this becomes
\begin{equation}
\begin{aligned}
    &U_0\left(\mathbf{c}^\dagger T^{1} \mathbf{c} - 2\right)^2 - U_K [\left(\mathbf{c}^\dagger T^{5} \mathbf{c}\right)^2 + \left(\mathbf{c}^\dagger T^{9} \mathbf{c}\right)^2] - U_N [\left(\mathbf{c}^\dagger T^{14} \mathbf{c}\right)^2 + \left(\mathbf{c}^\dagger T^{15} \mathbf{c}\right)^2 + \left(\mathbf{c}^\dagger T^{16} \mathbf{c}\right)^2] \\ & =  (U_0+U_N)\left(\mathbf{c}^\dagger T^{1} \mathbf{c} - 2\right)^2 - (U_K+U_N) \left(\mathbf{c}^\dagger T^{5} \mathbf{c}\right)^2 - (U_K+U_N) \left(\mathbf{c}^\dagger T^{9} \mathbf{c}\right)^2 + 2U_N \left(\mathbf{c}^\dagger T^{13} \mathbf{c}\right)^2.
\end{aligned}
\end{equation}
\end{widetext}
For notational clarity, we relabel $T^1$, $T^5$, $T^9$ and $T^{13}$ as $O^{0}$, $O^1$, $O^2$ and $O^3$, respectively. 
The transformed Hamiltonian, which is identical to $\mathcal{H}$, then takes the compact form 
\begin{equation}\label{eq:newmodel}
    H=\sum_{i=0}^{3} g_i H_i,
\end{equation}
where the coupling terms are $g_0 = U_0+U_N$, $g_1=-(U_K+N_N)$, $g_2=-(U_K+N_N)$ and $g_3=2U_N$ and operators are $H_i =\sum_{\mathbf{q}}^{N_{\mathbf{q}}} n^{O^{i}}_{\mathbf{q}} n^{O^{i}}_{-\mathbf{q}}$.

\subsection{Continuous Hubbard-Stratonovich transformation}
\label{sec:methods_hs}
To simulate the quartic fermionic Hamiltonian with DQMC, we use a continuous HS transformation to decouple the quartic interactions into quadratic fermionic terms coupled to auxiliary fields. We first perform a symmetric Trotter decomposition of the partition function $Z=\operatorname{Tr}[e^{-\beta H}]$, discretizing the inverse temperature $\beta$ into $N_\tau$ imaginary time slices of width $\Delta\tau=\beta/N_\tau$:
\begin{widetext}
\begin{equation}
  Z=\operatorname{Tr}\left[\left(e^{-\Delta\tau g_0 H_0/2} e^{-\Delta\tau g_1 H_1/2} e^{-\Delta\tau g_2 H_2/2} e^{-\Delta\tau g_3 H_3}  e^{-\Delta\tau g_2 H_2/2} e^{-\Delta\tau g_1 H_1/2} e^{-\Delta\tau g_0 H_0/2 } \right)^{N_\tau}\right]+\mathcal{O}\left(\Delta\tau^2\right),
  \label{eq:trotter_decomposition_methods}
\end{equation}
\end{widetext}
which has a systematic Trotter error of $O(\Delta\tau^2)$.

Next, we rewrite each interaction term $H_i$ as a sum of squares of Hermitian operators:
\begin{equation}
H_i=\sum_{\mathbf{q}} \frac{1}{2} \left[ (n^{O^{i}}_{\mathbf{q}}+n^{O^{i}}_{-\mathbf{q}})^2 - (n^{O^{i}}_{\mathbf{q}}-n^{O^{i}}_{-\mathbf{q}})^2\right] = \sum_{q=1}^{2N_q} \frac{(Q_q^i)^2}{2},
\label{eq:interaction_squares_methods}
\end{equation}
where $Q_q^i$ are Hermitian fermionic bilinear operators. We then apply the continuous HS transformation, based on the Gaussian integral identity:
\begin{equation}
e^{-\frac{\Delta\tau g_i }{2} (Q_q^i)^2}=\frac{1}{\sqrt{2\pi}}\int_{-\infty}^{\infty} d\phi_{q}^i e^{-\phi_{q}^{i^2}/2} e^{\alpha\phi_{q}^i Q_q^i},
\label{eq:hs_transform_methods}
\end{equation}
where $\alpha=\sqrt{-\Delta\tau g_i}$, and $\phi_q^i$ is a continuous real-valued auxiliary field.

Substituting Eq.~\eqref{eq:hs_transform_methods} into the Trotter-decomposed partition function, we obtain
\begin{equation}
Z = \int \mathcal{D}\phi \, e^{-S_0[\phi]} \operatorname{det}\left[1+B(\beta,0;\phi)\right],
\end{equation}
where $S_0[\phi]=\frac{1}{2}\sum_{\tau,i,q} (\phi_{\tau,q}^i)^2$ is the Gaussian action for the auxiliary fields, and $B(\beta,0;\phi)$ is the single-particle imaginary-time propagator constructed from the auxiliary field configuration $\phi$. The equal-time single-particle Green's function is given by
\begin{equation}
G(\tau,\tau;\phi) = \left(1+B(\tau,0;\phi)B(\beta,\tau;\phi)\right)^{-1},
\label{eq:green_function_methods}
\end{equation}
from which all physical observables can be computed via Wick's theorem.

\subsection{Langevin-based global update scheme}
\label{sec:methods_langevin}
For a given imaginary time slice $\tau$, we propose a new auxiliary field configuration $\phi_{\tau,q}^{'i}$ from the current configuration $\phi_{\tau,q}^i$ via the Langevin dynamics equation:
\begin{equation}
    \phi_{\tau,q}^{i\ \prime} = \phi_{\tau,q}^i + \frac{\epsilon^2}{2}  \nabla_{\phi_{\tau,q}^i} S[\phi] +\epsilon \mathcal{N}(0,1),
\label{eq:langevin_proposal_methods}
\end{equation}
where $\epsilon$ is the Langevin step size, $\mathcal{N}(0,1)$ is the standard normal distribution, and $S[\phi] = S_0[\phi] - \ln\operatorname{det}\left[1+B(\beta,0;\phi)\right]$ is the total effective action of the auxiliary fields.

The critical ingredient of this update is the analytical gradient of the action, which we derive using Jacobi's formula for the derivative of a determinant:
\begin{equation}
\nabla_{\phi_{\tau,q}^i} S[\phi] = \phi_{\tau,q}^i - \operatorname{Tr}\left[ G(\tau,\tau;\phi) \frac{\partial B(\tau,0;\phi)}{\partial \phi_{\tau,q}^i} \right].
\label{eq:action_gradient_methods}
\end{equation}
This gradient can be computed directly from the equal-time Green's function in Eq.~\eqref{eq:green_function_methods} with a computational complexity of $O(N_\phi^2)$ per auxiliary field.

The proposed configuration $\phi'$ is then accepted or rejected according to the Metropolis-Hastings criterion:
\begin{equation}
    P_{\text{acc}} = \min\left\{1, \frac{e^{-S[\phi']} f(\phi \mid \phi')}{e^{-S[\phi]} f(\phi' \mid \phi)} \right\},
\end{equation}
where $f(\phi' \mid \phi)$ is the proposal density from Eq.~\eqref{eq:langevin_proposal_methods}, given by  
\begin{equation}
    f(\mathbf{x} \mid \mathbf{y}) \propto \exp\left[-\frac{1}{2\epsilon^2} \left\| \mathbf{y} - \mathbf{x} - \frac{\epsilon^2}{2} \nabla S(\mathbf{x}) \right\|_2^2\right],
\end{equation}
with $\|\cdot\|_2$ denoting the standard $L^2$-norm.

To optimize the update efficiency, we implement an adaptive step size adjustment mechanism. The Langevin step size $\epsilon$ is automatically tuned during the simulation to maintain an average acceptance rate of $r_0=0.574$\cite{robertsOptimal1998,robertsOptimal2001}, via the update rule:
\begin{equation}
\epsilon \rightarrow \epsilon + (r_a - r_0) \Delta\epsilon,
\end{equation}
where $r_a$ is the average acceptance rate over the previous sweep, and $\Delta\epsilon=0.1$ is a small tuning parameter.

\subsection{Benchmark against density matrix renormalization group calculations}
\label{sec:methods_benchmark}
\begin{figure}[htp!]
\includegraphics[width=\columnwidth]{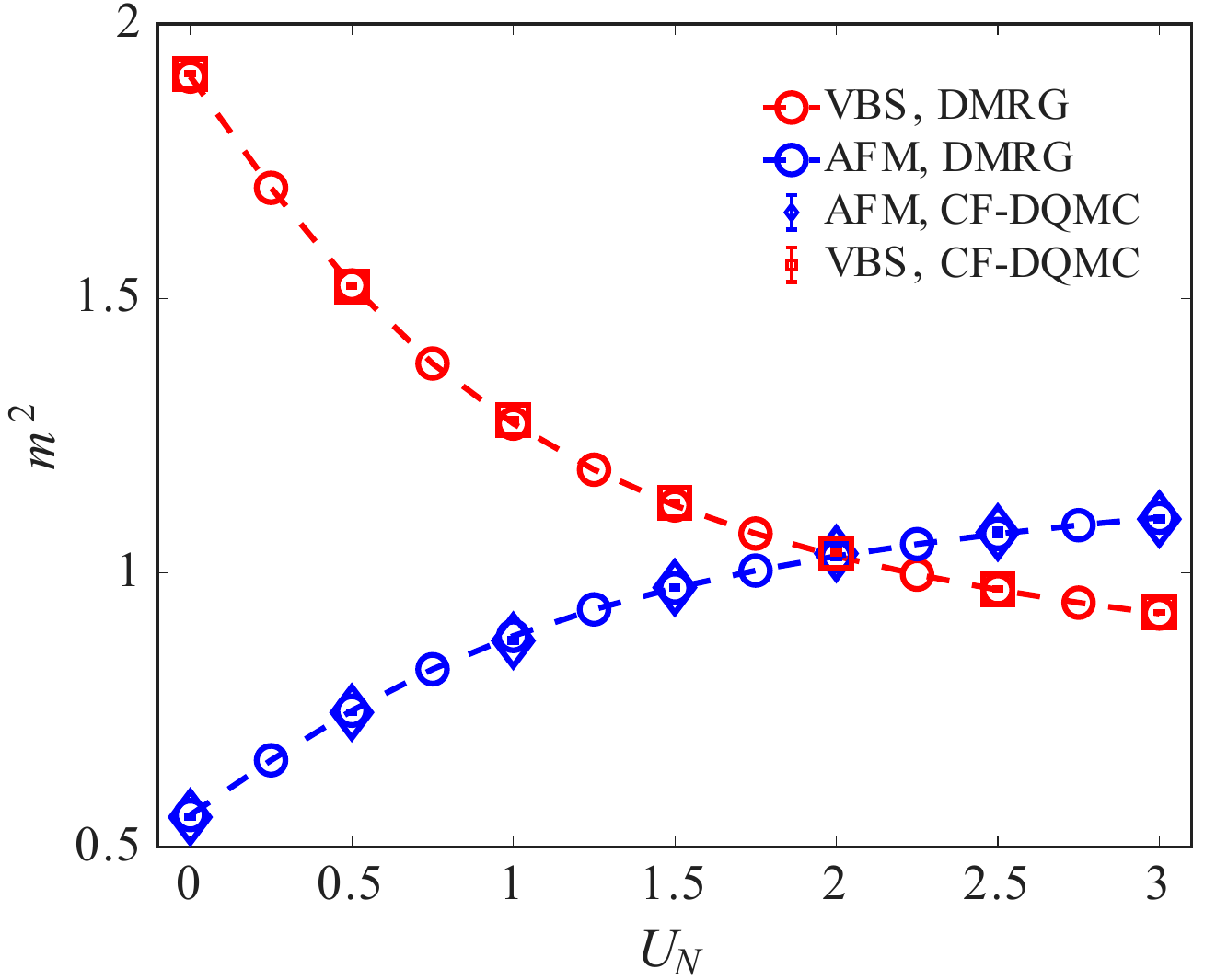}
\caption{\textbf{Benchmark of the CF-DQMC algorithm against DMRG.} VBS order parameter $m_{\text{VBS}}^2$ as a function of $U_N$ for fixed $U_K=2$, $N_\phi=7$, $\beta=8N_\phi$, and $\Delta\tau=0.01$ on the spherical geometry. Our CF-DQMC results (blue squares) are in excellent agreement with DMRG results (red circles) from Ref.~\cite{chenPhases2023}, confirming the accuracy and unbiased nature of our algorithm.}
\label{fig:benchmark_methods}
\end{figure}

To validate our CF-DQMC algorithm, we have performed benchmark simulations on the spherical geometry for small system sizes, and compared our results to DMRG calculations from Ref.~\cite{chenPhases2023}. Fig.~\ref{fig:benchmark_methods} shows the VBS order parameter $m_{\text{VBS}}^2$ as a function of $U_N$ for fixed $U_K=2$ and $N_\phi=7$, computed with our CF-DQMC algorithm and with DMRG. The results are in excellent agreement across the entire parameter range, confirming that our algorithm is unbiased and accurate.

\section*{Data availability}
All data that support the findings of this study are
available from the corresponding authors upon request. Source data
are provided within this paper.

\section*{Code availability}
The codes used in this study are available from the corresponding
author upon request.

\bibliographystyle{shortapsrev4-2}
\bibliography{bibtex}

\section*{Acknowledgements}
We thank Zhenjiu Wang, Maksim Ulybyshev, David Poland, Cenke Xu, Meng Cheng, Yi-Zhuang You, and Matthias Vojta for helpful discussions.
YDL acknowledges support from the National Natural Science Foundation of China (Grant No. 12404282).
YDL and ZYM acknowledge support from General Program of the Guangdong Natural Science Foundation (Grant No. 2025A1515010337), the Research Grants Council (RGC) of Hong Kong (Project Nos. AoE/P-701/20,  C7037-22GF, 17302223, 17301924, 17301725), the ANR/RGC Joint Research Scheme sponsored by RGC of Hong Kong and French National Research Agency (Project No. A\_HKU703/22) and the State Key Laboratory of Optical Quantum Materials at HKU. 
LJ acknowledges support by the Deutsche Forschungsgemeinschaft (DFG) through SFB 1143 (A07, Project No.~247310070), the W\"urzburg-Dresden Cluster of Excellence {\it ctd.qmat} (EXC 2147, Project No.~390858490), and the Emmy Noether program (JA2306/4-1, Project No.~411750675).
FFA acknowledges support by  the Deutsche Forschungsgemeinschaft  through AS 120/19-1 (Project No.~530989922) and the W\"urzburg-Dresden Cluster of Excellence {\it ctd.qmat} (EXC 2147, Project No.~390858490)
JR is funded by the European Union (ERC, QFTinAdS, project number 101087025, PI Balt van Rees).
Views and opinions expressed are, however, those of the authors only and do not necessarily reflect those of the European Union or the European Research Council Executive Agency.
We thank HPC2021 system under the Information Technology Services at the University of Hong Kong~\cite{hpc2021}, as well as the Beijing Paratera Tech Corp., Ltd~\cite{paratera} for providing HPC resources that have contributed to the research results reported within this paper. We further gratefully acknowledge the computing time made available on the high-performance computer Barnard at the NHR Center of TU Dresden. This center is jointly supported by the Federal Ministry of Education and Research and the state governments participating in the National High-Performance Computing (NHR) joint funding program~\cite{nhr}.

\newpage

\clearpage
\onecolumngrid

\begin{center}
\begin{spacing}{1.5}
\textbf{\Large Supplementary Information for\\ ``Numerical evidence of a critical point in the (2+1)D SO(5) nonlinear sigma model with Wess-Zumino-Witten term''}
\end{spacing}
\end{center}
\setcounter{equation}{0}
\setcounter{figure}{0}
\setcounter{table}{0}
\setcounter{page}{1}
\setcounter{section}{0}
  
\makeatletter
\renewcommand{\theequation}{S\arabic{equation}}
\renewcommand{\thefigure}{S\arabic{figure}}

\section*{Supplementary Note 1: Phase boundaries on the toric geometry away from the SO(5)-symmetric line}
\label{sec:supp_phase_boundaries_torus}

\begin{figure}[tb!]
\includegraphics[width=0.66\columnwidth]{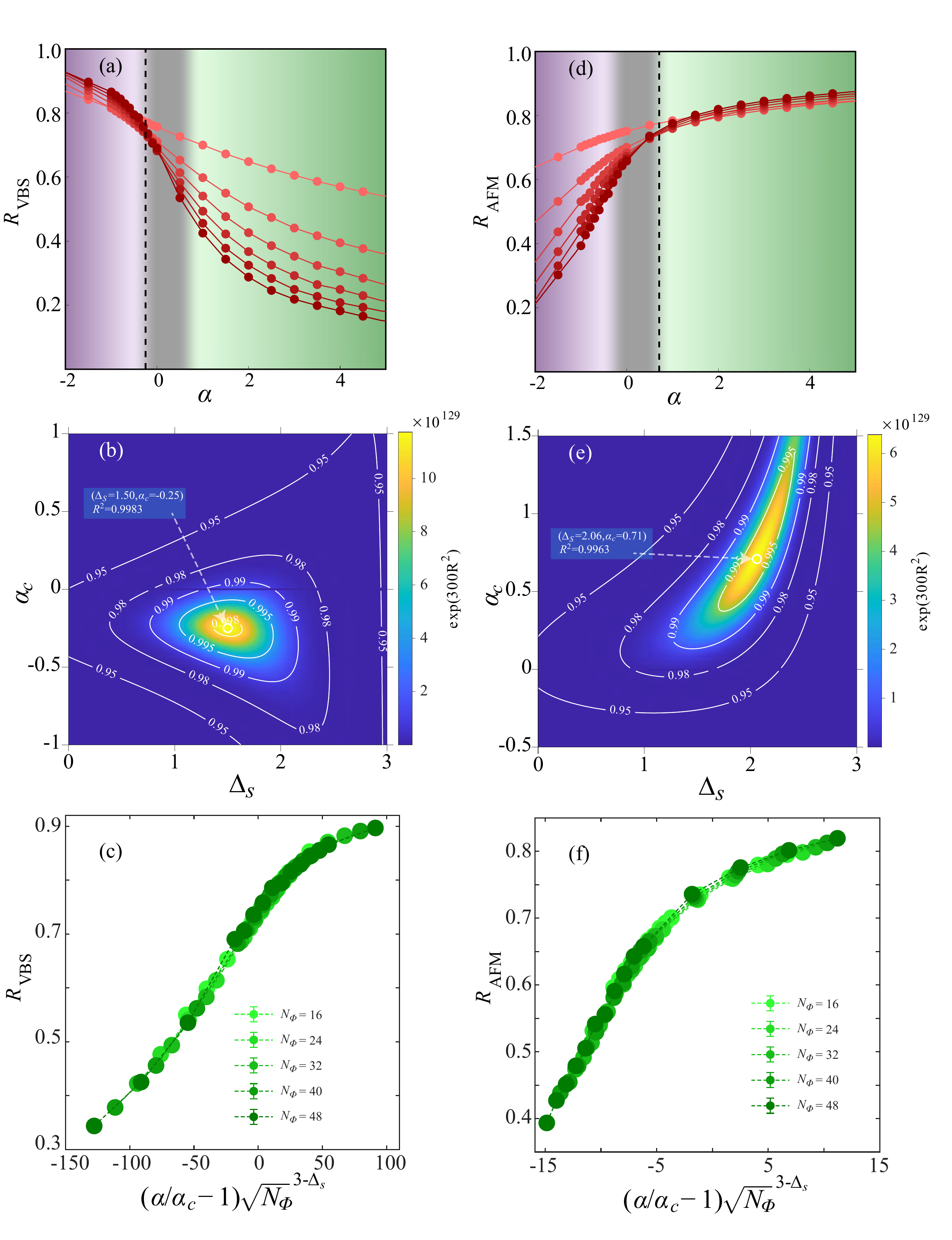}
\caption{\textbf{Phase boundaries and finite-size scaling analysis away from the SO(5)-symmetric line on the torus geometry at fixed $u=3$. } (a) VBS correlation ratio $R_{\text{VBS}}$ as a function of anisotropy $\alpha$ for increasing system sizes, with cyan, gray, and green backgrounds marking the VBS, disordered, and AFM phases, respectively. Curve crossings identify the VBS–disorder transition. (b) $R^2$ goodness-of-fit map from grid scanning of $R_{\text{VBS}}$. (c) Data collapse of $R_{\text{VBS}}$ obtained with the optimal critical parameters. (d)--(f) Corresponding analysis for the disorder--AFM transition: (d) AFM correlation ratio versus $\alpha$, (e) $R^2$ goodness-of-fit map, and (f) data collapse of the AFM correlation ratio. Panels (a) and (d) display data for system sizes $N_\phi = 8, 16, 24, 32, 40, 48$, with curves colored from light to dark as $N_\phi$ increases.}
\label{fig:awaySO5_torus}
\end{figure}

We perform a parallel finite-size scaling analysis of the VBS–disorder and disorder–AFM phase transitions on the torus geometry, adopting the same $R^2$-optimized grid scanning and data collapse protocol used for the sphere in the main text. All results are computed at fixed coupling $u=3$, deep within the SO(5)-symmetric disordered phase, while the anisotropy parameter $\alpha$ is tuned to traverse both phase transitions. We consider system sizes $N_\phi = 8, 16, 24, 32, 40, 48$, and present the full set of results in Fig.~\ref{fig:awaySO5_torus}.

Panels (a)–(c) of Fig.~\ref{fig:awaySO5_torus} detail the VBS–disorder transition. The VBS correlation ratio $R_{\text{VBS}}$ exhibits well-defined crossing points across different system sizes at a finite negative value of $\alpha$, confirming that the transition occurs at a finite offset from the SO(5)-symmetric line. We carry out an $R^2$-optimized grid scan, restricting the scaling analysis to system sizes $N_\phi \ge 16$ and approximating the universal scaling function $\mathcal{F}(x)$ by a third-order polynomial. The best-fit critical parameters are found to be $\alpha_{c,\text{VBS}} = -0.25$ and $\Delta_{S,\text{VBS}} = 1.50$. The corresponding data collapse, shown in panel (c), is of good quality, consistent with a continuous quantum phase transition between the VBS phase and the SO(5)-symmetric disordered phase.

Panels (d)–(f) present the analogous analysis for the disorder–AFM transition. Applying the same crossing-point identification and $R^2$-optimized grid scanning procedure, we obtain best-fit critical parameters $\alpha_{c,\text{AFM}} = 0.71$ and $\Delta_{S,\text{AFM}} = 2.06$. The data collapse displayed in panel (f) further corroborates the continuous nature of the disorder–AFM transition.

Overall, the critical parameters extracted on the torus are in good quantitative agreement with those reported for the sphere geometry in the main text. 

\section*{Supplementary Note 2: Dynamical critical exponent measurement in the disordered phase}
To characterize the low-energy dynamics of the extended SO(5) symmetric disordered phase, we compute the dynamical critical exponent $z$ via finite-size scaling of the imaginary-time SO(5) correlation function. The measurement is performed deep in the disordered phase at $u=15$, far from the multicritical point at $u\approx0.2$, for both spherical and toric geometries.

For each system size $N_\phi$, we first compute the imaginary-time correlation function of the SO(5) order parameter at $\Gamma$ point:
\begin{equation}
m^2_{\text{SO(5)}}(\tau, \mathbf{q}=0) = \frac{1}{5}\sum_{i=1}^5 \langle n^{\Gamma^i}_{\mathbf{q}=0}(\tau) n^{\Gamma^i}_{\mathbf{q}=0}(0) \rangle,
\end{equation}
where $n^{\Gamma^i}_{\mathbf{q}}(\tau)$ is the Fourier-transformed density operator for the $i$-th SO(5) component at imaginary time $\tau$. At long imaginary times, this correlation function decays exponentially as $m^2_{\text{SO(5)}}(\tau, 0) \sim e^{-\Delta_{\text{SO5}} \tau}$, where $\Delta_{\text{SO5}}$ is the finite-size gap of the SO(5) vector excitation. We extract $\Delta_{\text{SO5}}$ for each system size by fitting the long-time tail of the correlation function to this exponential form.

For a quantum critical system with linear size $L = \sqrt{N_\phi}$, the finite-size gap scales with system size as $\Delta_{\text{SO5}} \sim L^{-z} = (\sqrt{N_\phi})^{-z}$, where $z$ is the dynamical critical exponent. Taking the natural logarithm of both sides gives the linear relation:
\begin{equation}
\ln\Delta_{\text{SO5}} = z \cdot \ln\left(\frac{1}{\sqrt{N_\phi}}\right) + C,
\end{equation}
where $C$ is a non-universal constant. The slope of the linear fit between $\ln\Delta_{\text{SO5}}$ and $\ln(1/\sqrt{N_\phi})$ thus directly yields the dynamical exponent $z$.

\begin{figure}[htp!]
\includegraphics[width=\columnwidth]{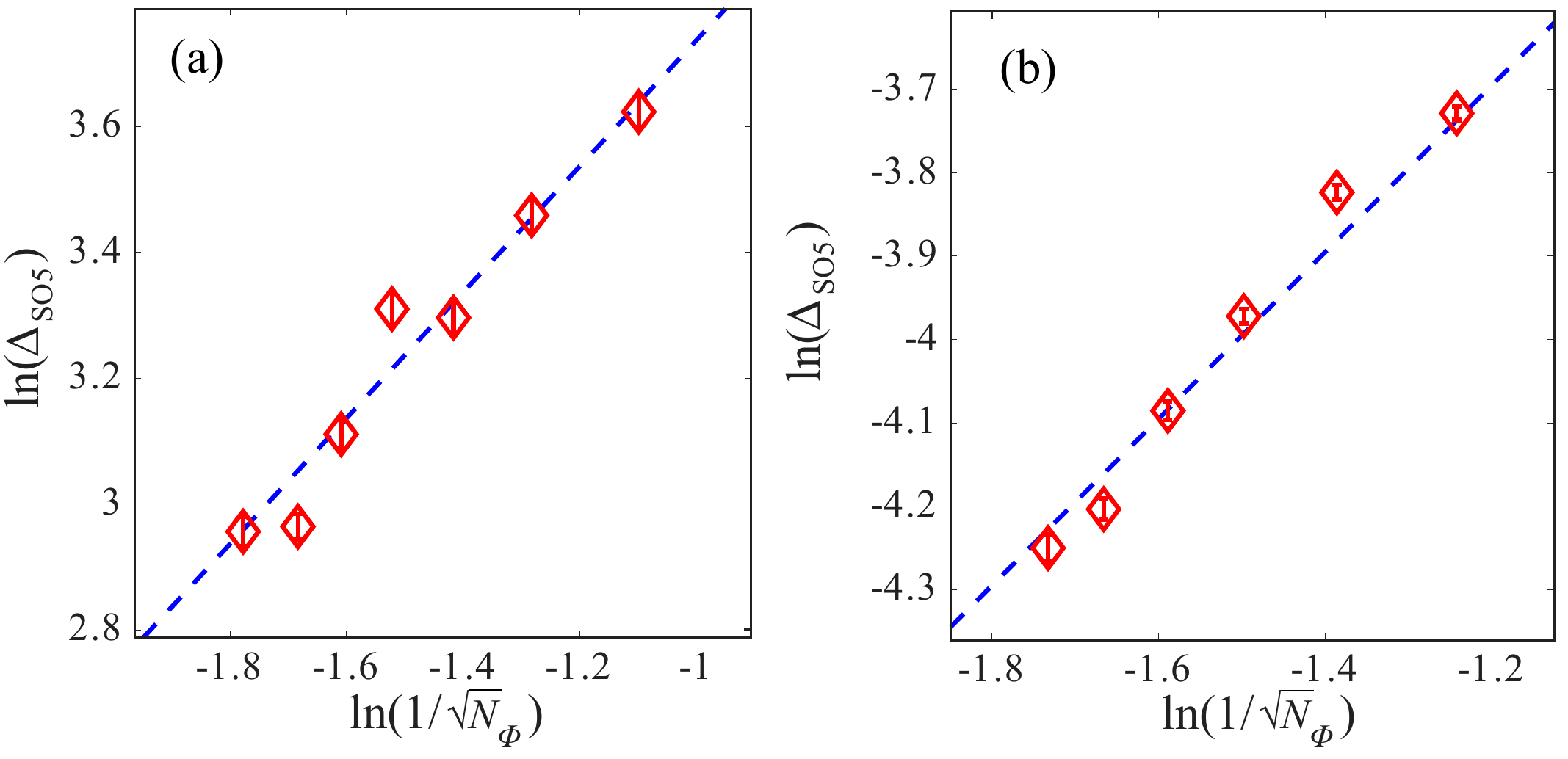}
\caption{\textbf{Finite-size scaling of the SO(5) gap for dynamical critical exponent measurement.}
Log-log plot of the extracted SO(5) gap $\Delta_{\text{SO5}}$ versus inverse linear system size $1/\sqrt{N_\phi}$ for the disordered phase at $u=15$.
(a) Results for the spherical geometry, with system sizes up to $N_\phi=35$.
(b) Results for the toric geometry, with system sizes up to $N_\phi=32$.
In both panels, red diamonds represent the measured gap values from exponential fits to the imaginary-time SO(5) correlation function, and the blue dashed line shows the relation $\ln\Delta_{\text{SO5}} = z \cdot \ln\left({1}/{\sqrt{N_\phi}}\right) + C$ with slope $z=1$, corresponding to a Lorentz-invariant theory. }
\label{fig:zvalue}
\end{figure}

Our results for this scaling analysis are presented in Fig.~\ref{fig:zvalue}. For both geometries, the measured data points follow a clear linear trend, with slopes roughly consistent with $z=1$ within statistical error bars.

\section*{Supplementary Note 3: Finite-size scaling analysis on the toric geometry and correlation ratio scaling}
\label{app:torus_scaling}

In this section, we present the finite-size scaling analysis of the SO(5) order parameter on the toric geometry, as well as a complementary scaling analysis of the correlation ratio for both spherical and toric geometries. These results corroborate the universal phase structure and critical behavior identified in the main text.

\subsection{Finite-size scaling on the toric geometry}
\label{app:torus_fss}

\begin{figure}[tb!]
\includegraphics[width=0.6\columnwidth]{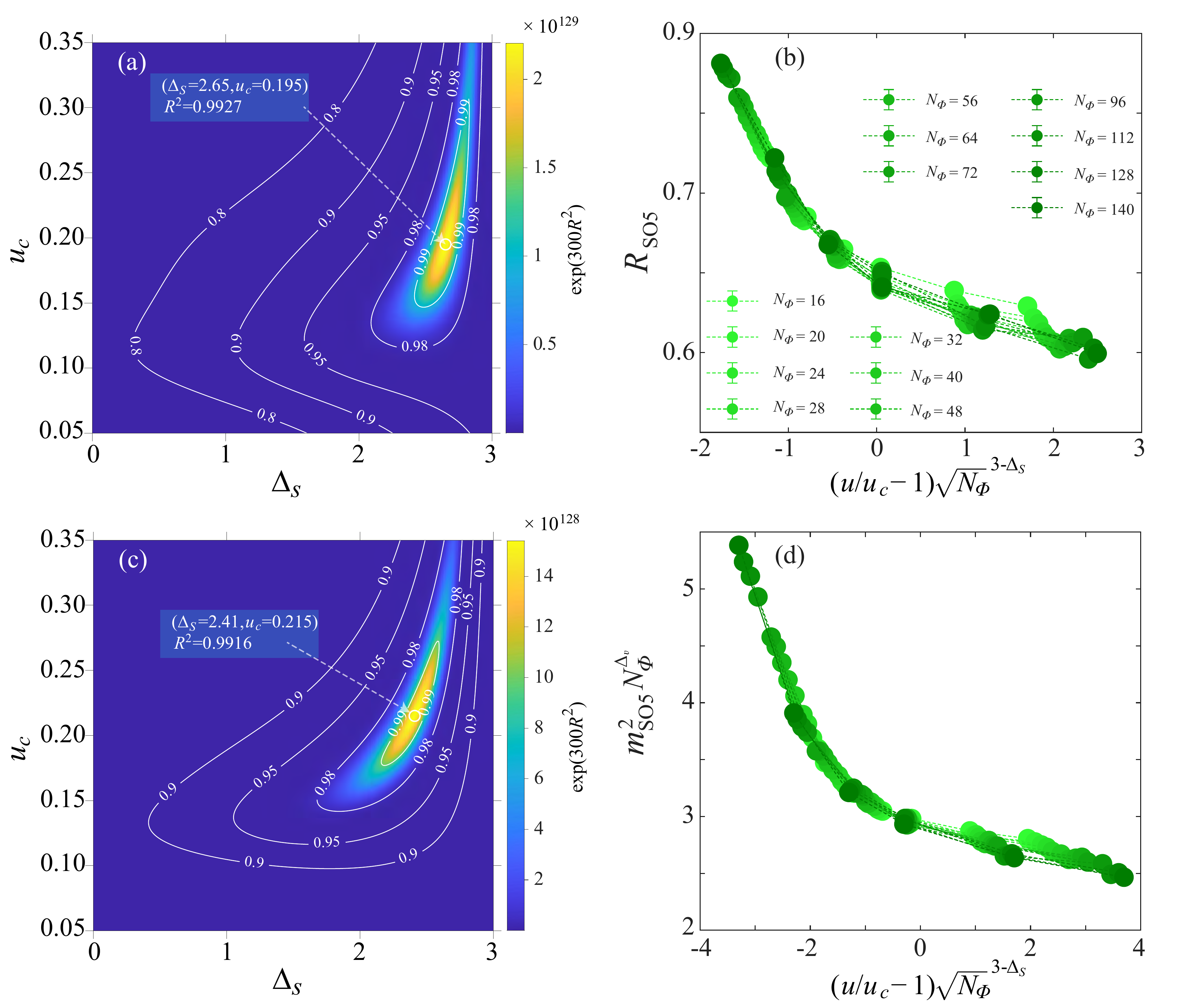}
\caption{\textbf{Finite-size scaling analysis along the SO(5)-symmetric line on the torus geometry.} (a) $R^2$ goodness-of-fit map from grid scanning of the correlation ratio $R_{\text{SO(5)}}$ versus critical coupling $u_c$ and scaling dimension $\Delta_s$. (b) Data collapse of $R_{\text{SO(5)}}$ for $N_\phi \ge 16$, obtained with optimal parameters $u_c=0.195$, $\Delta_s=2.65$. (c) $R^2$ goodness-of-fit map from grid scanning of the squared SO(5) order parameter $m_{\text{SO(5)}}^2$ at fixed $\Delta_\varv=0.51$. (d) Data collapse of $m_{\text{SO(5)}}^2$ using the optimal parameters $u_c=0.215$, $\Delta_s=2.41$ extracted from panel (c).
}
\label{fig:cross-collapse-order-R2-torus}
\end{figure}

We carry out a parallel finite-size scaling analysis along the SO(5)-symmetric line on the torus geometry, adopting the identical $R^2$-optimized grid scanning and data collapse protocol described for the sphere geometry in the main text.

For the toric lattice, we investigate system sizes $N_\phi \in \{16, 20, 24, 28, 32, 40, 48, 56, 64, 72, 96, 112, 128, 140\}$, with the largest system reaching $N_\phi=140$, and we sample couplings over the extended range $u \in [0.05, 0.4]$.

We begin with the correlation ratio $R_{\text{SO(5)}}$, which satisfies the same finite-size scaling ansatz as in the main text:
$$
R = \mathcal{F}\left( (u/u_{c}-1) \sqrt{N_\phi}^{3-\Delta_s} \right),
$$
where $\mathcal{F}(x)$ is the universal scaling function. Approximating $\mathcal{F}(x)$ as a third-order polynomial in the critical region, we perform a grid scan over $u_c$ and $\Delta_s$ to optimize the $R^2$ coefficient of the data collapse. The resulting goodness-of-fit map is shown in Fig.~\ref{fig:cross-collapse-order-R2-torus}(a), from which we obtain the best-fit parameters $u_{c}=0.195$ and $\Delta_s=2.65$. This critical coupling $u_c$ is in full agreement with the transition location inferred from the system-size dependence of the SO(5) order parameter on the torus, presented in Fig.~2(d) of the main text.

The corresponding data collapse of $R_{\text{SO(5)}}$ for all $N_\phi \ge 16$ is plotted in Fig.~\ref{fig:cross-collapse-order-R2-torus}(b). Although the collapse quality is moderately lower than that observed on the sphere geometry, all data points collapse onto a single universal curve, confirming the critical behavior of the transition.

We repeat the scaling analysis for the squared SO(5) order parameter $m_{\text{SO(5)}}^2$, using the scaling form
$$
m_{\text{SO(5)}}^2 N_\phi^{\Delta_\varv} = \mathcal{F}\left( (u/u_{c}-1) \sqrt{N_\phi}^{3-\Delta_s} \right).
$$
Fixing the order-parameter scaling dimension to $\Delta_\varv=0.51$, consistent with the value extracted from Fig.~2(d) of the main text, we obtain the optimal fit $u_{c}=0.215$ and $\Delta_s=2.41$, as illustrated in the $R^2$ map of Fig.~\ref{fig:cross-collapse-order-R2-torus}(c).

The associated data collapse is shown in Fig.~\ref{fig:cross-collapse-order-R2-torus}(d). As with the correlation ratio, the collapse for the order parameter is of good quality, albeit slightly less sharp than on the sphere, which we attribute to the stronger finite-size corrections inherent to periodic boundary conditions on the torus.

In summary, the critical parameters extracted independently from $R_{\text{SO(5)}}$ and $m_{\text{SO(5)}}^2$ on the torus are mutually consistent within error, and also agree with the sphere results reported in the main text. The toric analysis therefore provides independent confirmation of the continuous quantum phase transition between the SO(5)-ordered and SO(5)-disordered phases, and lends further support to the estimated critical exponents.

\subsection{Data Scaling Analysis}
\label{app:data_crosing}

\begin{figure}[tb!]
\includegraphics[width=\columnwidth]{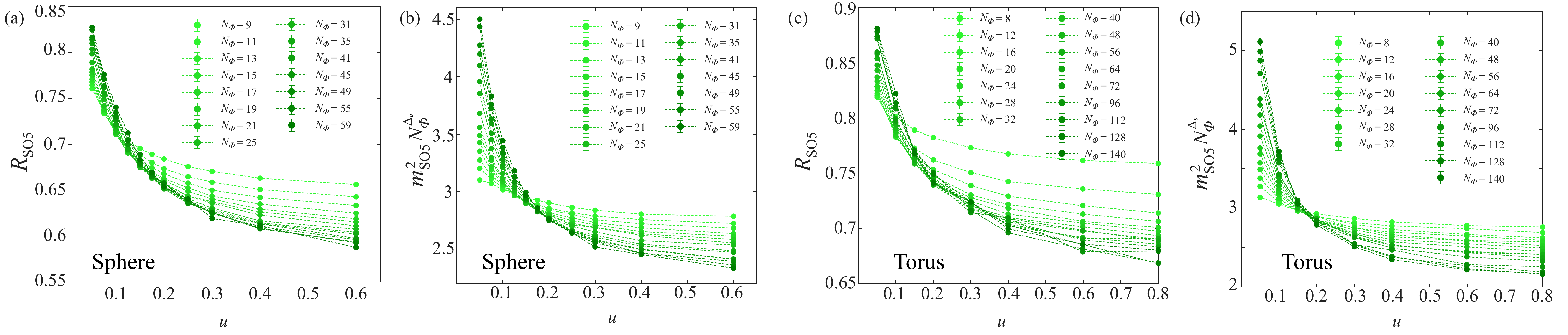}
\caption{\textbf{Data crossing analysis of the SO(5) quantum phase transition on the sphere and torus geometries.} (a) Correlation ratio $R_{\text{SO(5)}}$ as a function of coupling $u$ for increasing system sizes on the sphere, where the common crossing of curves marks the critical point. (b) Scaled squared SO(5) order parameter $m_{\text{SO(5)}}^2 N_\phi^{\Delta_\varv}$ versus $u$ on the sphere. (c), (d) Corresponding results for the correlation ratio and scaled order parameter on the torus geometry.
}
\label{fig:data_crosing}
\end{figure}

Figure~\ref{fig:data_crosing} compiles the data crossing results for both geometries. On the spherical geometry [panels (a) and (b)], the correlation ratio curves for different system sizes intersect at $u_\mathrm{c} \approx 0.175$, and the scaled squared SO(5) order parameter $m_{\text{SO(5)}}^2 N_\phi^{\Delta_\varv}$ yields a consistent crossing at the same coupling. On the toric geometry [panels (c) and (d)], both the correlation ratio and the scaled order parameter exhibit a common crossing point at $u_\mathrm{c} \approx 0.20$.

This cross-validation across two independent observables and two distinct lattice geometries lends further robust support to our conclusion that the SO(5) nonlinear sigma model with a Wess-Zumino-Witten term hosts a continuous quantum phase transition characterized by universal critical behavior.

\end{document}